\documentclass[11pt,a4paper]{article}
\usepackage{jheppub}
\pdfoutput=1

\hypersetup{citecolor=black,linkcolor=black}

\usepackage[usenames, dvipsnames]{xcolor}

\usepackage{booktabs}
\usepackage{multirow}
\usepackage{amssymb}
\usepackage[export]{adjustbox}
\usepackage{floatrow}
\newfloatcommand{capbtabbox}{table}[][3.5in]
\newfloatcommand{capbfigbox}{figure}[][2.3in]
\usepackage{blindtext}
\usepackage{amsmath}
\usepackage{booktabs}
\usepackage{nicefrac}

\graphicspath{{Figures/}}

\newcommand{\be}{\begin{equation}}
\newcommand{\ee}{\end{equation}}
\newcommand{\nl}{\nonumber \\}

\newcommand{\ie}{i.e.~}
\newcommand{\eg}{e.g.~}

\def\lsim{\mathrel{\raise.3ex\hbox{$<$\kern-.75em\lower1ex\hbox{$\sim$}}}}
\def\gsim{\mathrel{\raise.3ex\hbox{$>$\kern-.75em\lower1ex\hbox{$\sim$}}}}

\begin{document}

\preprint{FERMILAB-PUB-16-456-T\\ \hspace*{114mm}{UCI-HEP-TR-2016-19}}


\title{Multi-component Dark Matter through a Radiative Higgs Portal}

\author[a,b,c]{Anthony DiFranzo}
\emailAdd{adifranz@uci.edu}

\author[a,d]{and Gopolang Mohlabeng}
\emailAdd{gopolang.mohlabeng@ku.edu}

\affiliation[a]{Theoretical Physics Department, Fermilab, Batavia, IL 60510, USA}
\affiliation[b]{Department of Physics and Astronomy, University of California, Irvine, CA 92697, USA}
\affiliation[c]{Department of Physics and Astronomy, Rutgers University, Piscataway, NJ 08854, USA}
\affiliation[d]{Department of Physics and Astronomy, University of Kansas, Lawrence, KS 66045, USA}

\abstract{We study a multi-component dark matter model where interactions with the Standard Model are primarily via the Higgs boson. The model contains vector-like fermions charged under $SU(2)_W \times U(1)_Y$ and under the dark gauge group, $U(1)^\prime$. This results in two dark matter candidates. A spin-1 and a spin-\nicefrac{1}{2} candidate, which have loop and tree-level couplings to the Higgs, respectively. We explore the resulting effect on the dark matter relic abundance, while also evaluating constraints on the Higgs invisible width and from direct detection experiments. Generally, we find that this model is highly constrained when the fermionic candidate is the predominant fraction of the dark matter relic abundance.}

\maketitle


\section{Introduction}
\label{sec:intro}
Dark Matter (DM) remains one of the most profound mysteries in Nature. Its existence has been very well established by an overwhelming amount of astrophysical data. We know very little about its microscopic nature \cite{Arrenberg:2013rzp}. Even so, unfortunately, the Standard Model (SM) of particle physics cannot accommodate DM, making it essential to consider a balanced discovery effort based on various DM searches.

Among myriad possibilities, scenarios with multiple dark matter candidates are very well motivated and have been investigated from the largest scales in Cosmology to the smallest scales at experiments on earth \cite{Semenov:2013vya, Peim:2014zia, Konar:2009qr}. Furthermore, multicomponent DM can provide an alternative solution to the small scale structure problems that are a result of the discrepancy between collisionless cold dark matter and observational data \cite{Medvedev:2013vsa}. In addition, studies with multiple dark matter particles have illustrated very interesting phenomenology as well as the potential of providing ``smoking gun" signatures in both particle and astrophysical experiments \cite{Agashe:2014yua, Kong:2014mia, Kopp:2015bfa, Berger:2014sqa, Dienes:2011ja, Dienes:2014via, Aoki:2013gzs, Aoki:2014lha, Geng:2013nda, Geng:2014dea, Peim:2014zia, Bian:2013wna, Barger:2008jx, Bian:2014cja, Duch:2015jta, Bhattacharya:2013hva, Bhattacharya:2013asa, Berezhiani:1990sy, Chialva:2012rq}.

On the other hand, the SM Higgs boson has a very important role to play in nature. As the facilitator of electroweak symmetry breaking (EWSB), it can provide a window into new weak-scale physics beyond the SM. In particular, as the only elementary scalar in the SM, it can be the means through which new physics communicates with the SM, this can occur through the gauge invariant, low dimensional bilinear operator, $H^{\dagger} H$. As a consequence, fundamental questions such as the naturalness problem and the Higgs vacuum stability may be addressed by introducing new bosons which interact with the Higgs bilinear operator. These interactions can occur through a vector or scalar Higgs portal in the form of $\lambda H^{\dagger}H \Phi^{\dagger} \Phi$ and $\lambda H^{\dagger}H V^{\mu} V_{\mu}$ respectively, with $\lambda$ being some dimensionless coupling. A further possibility is to have interactions of the Higgs with new heavy fermions through higher dimensional operators, a fermion Higgs portal. Any of these new particles can, in principle, constitute DM allowing for a direct glimpse into the dark sector.

In this paper, we consider the model outlined in Ref.~\cite{DiFranzo:2015nli}. The phenomenology consists of a gauged, dark $U(1)^{\prime}$ symmetry. The corresponding gauge boson, $V$, obtains a mass when the $U(1)^{\prime}$ symmetry is spontaneously broken by a SM singlet scalar, $\Phi$. Fermions charged under this $U(1)^{\prime}$ and the SM Electroweak groups, are also introduced. A dark charge conjugation symmetry is imposed, which must not be broken when $\Phi$ receives a vev, so as to ensure stability of the vector. However, this requires the lightest of the new fermions to also be stable. Direct coupling of the vector to the Higgs is forbidden, which results in its interaction with the SM only at the radiative level. 

This work is organized as follows. In Sec.~\ref{sec:radhiggs}, we provide a summary of the UV completion presented in Ref.~\cite{DiFranzo:2015nli} and expand on the stability mechanism. In Sec.~\ref{sec:abundance} we discuss the evolution of the number density of the DM species considering different phenomenological scenarios. We evaluate the thermal relic density, the direct detection cross-section, and the invisible Higgs width in Sec.~\ref{sec:pheno}. This is followed by a discussion of how these observables constrain the model in Sec.~\ref{sec:results}. Finally, we conclude in Sec.~\ref{sec:summary}.


\section{The Radiative Higgs model for Two Component DM}
\label{sec:radhiggs}
When writing down a UV completion to the typical Vector Higgs Portal ($H^{\dagger}H V^{\mu} V_{\mu}$), there are two previously proposed options. Both possibilities introduce a new gauge group, which when spontaneously broken generates a spin-1 dark matter candidate. The first portal is through mixing between the SM Higgs and the scalar which breaks the dark gauge group, resulting in a tree-level, mixing suppressed coupling between the Higgs and the vector \cite{Hambye:2008bq, Farzan:2012hh, Baek:2012se, Baek:2013qwa, Baek:2014jga, Baek:2014goa, Ko:2014gha, Gross:2015cwa, DiChiara:2015bua, Chen:2015dea, Kim:2015hda, Karam:2015jta, Karam:2016rsz}. The second option, which is of interest in this work, further introduces new fermions which carry dark and SM Electroweak charges. These fermions generate a loop-level coupling between the Higgs and vector \cite{DiFranzo:2015nli}.

The model explored in Ref.~\cite{DiFranzo:2015nli} proposes a $U(1)^\prime$ whose gauge field is denoted as $V$. The model contains matter which is anomaly free and does not induce a kinetic mixing between the dark and SM gauge bosons. This is detailed in Sec.~II of that work, which we summarize below.

The matter content of the model is given in Table \ref{tab:NPtran} with the following mass and Higgs interaction terms for the fermions:
\begin{equation}
\begin{aligned}
\mathcal L &~ \supset -m ~ \epsilon^{ab}  \left( \psi_{1a} \chi_{1b} + \psi_{2a} \chi_{2b} \right) - m_n ~ n_1 n_2\\
          &- y_{\psi}~\epsilon^{ab}  \left( \psi_{1a} H_b n_1 + \psi_{2a} H_b n_2 \right) - y_{\chi} \left( \chi_1 H^* n_2 + \chi_2 H^* n_1 \right) + h.c.
\end{aligned}
\label{eq:mass}
\end{equation}
\begin{table}
\centering
\caption{Charge assignments for ($\nicefrac{1}{2}$,0) Weyl fermions $\psi$, $\chi$, and $n$ and complex scalar $\Phi$.}
\begin{tabular}{cccc}
  ~~Field~~~ & ~~~($SU(2)_W$, $U(1)_Y$, $U(1)^\prime$)~~~~~~~~~~~~~
  & ~~Field~~~ & ~~~($SU(2)_W$, $U(1)_Y$, $U(1)^\prime$)
    \\
  \hline \hline                   
  $\psi_{1\alpha}$ & (2, \nicefrac{1}{2}, ~1)  &
  $\psi_{2\alpha}$ & (2, \nicefrac{1}{2}, -1)   \\
  $\chi_{1\alpha}$ & (2, \nicefrac{-1}{2}, -1) &
  $\chi_{2\alpha}$ & (2, \nicefrac{-1}{2}, ~1)   \\
  $n_{1\alpha}$ & (1, ~0, ~-1) &
  $n_{2\alpha}$ & (1, ~0, ~~1)  \\
  \hline
  $\Phi$~ & (1, ~0, ~~$Q_\Phi$) \\
\end{tabular}
\label{tab:NPtran}
\end{table}

In writing down this model, a $U(1)^\prime$ charge conjugation (CC$^\prime$) symmetry is imposed and whose transformation is given by the following prescription:
\begin{align}
f_1 &\longleftrightarrow f_2 \nonumber\\
V &\longrightarrow -V \\
\Phi &\longrightarrow \Phi^*\nonumber
\end{align}
Where $f$ stands for the $\psi$, $\chi$, and $n$ fermions.

Imposing CC$^\prime$ removes the tree-level kinetic mixing term between hypercharge and $U(1)^\prime$, $F^{\mu\nu}F^{\prime}_{\mu\nu}$, and aligns various yukawa couplings and masses appearing in Eq.~\ref{eq:mass}. Since we assume that $Q_\Phi\neq\pm1$ and that the Higgs is not charged under the $U(1)^\prime$, neither EWSB nor the spontaneous breaking of $U(1)^\prime$ lead to CC$^\prime$ violating terms.

One may be concerned that $\Phi$ spontaneously breaks CC$^\prime$. One is free to rotate $\Phi$ using the global $U(1)^\prime$, such that only the real component of $\Phi$ receives a vacuum expectation value. Under CC$^\prime$, $\Phi$ transforms as ${\rm Im}(\Phi)\rightarrow-{\rm Im}(\Phi)$, therefore CC$^\prime$ is left intact after $U(1)^\prime$ is broken\footnote{Alternatively, this may equivalently be seen without rotating $\Phi$. For general $\theta={\rm Arg}(\left<\Phi\right>)$, both CC$^\prime$ and the global $U(1)^\prime$ break. However, the subgroup whose transformation is $\Phi\rightarrow e^{2i\theta}\Phi^*$ is preserved. This would be identified as the new CC$^\prime$ symmetry.}. Note that the imaginary component of $\Phi$, being the $U(1)^\prime$ Goldstone boson, has the same transformations properties as $V$ under CC$^\prime$.

All perturbative processes which could break CC$^\prime$ rely on a tree-level source of breaking. Therefore, with these assumptions, once this symmetry has been imposed at tree-level, it is preserved at every order in perturbation theory. $V$ is odd under this symmetry, thus it can only decay to the new fermions. More precisely, if the fermions are heavy, \ie $2M_f>M_V$, $V$ is stable. This is in direct analogy to Furry's theorem of QED \cite{Furry:1937}.

However, note that CC$^\prime$ also forbids amplitudes with only one new fermion appearing in external lines. As pointed out in Ref.~\cite{DiFranzo:2015nli}, the lightest new fermion is also stable and, therefore, another dark matter candidate.

Previous work on this model restricted itself to regimes where the fermions were heavy. In this work, we wish to explore the regime where one fermion is light enough to be a relevant degree-of-freedom in dark matter phenomenology. From the perspective of relic abundance, there are two effects which motivate investigating this case. First, the vector candidate annihilates more efficiently for lighter fermions, since the annihilation rate is suppressed by the mass of the fermion. Further, when the fermion running in the $h$-$V$-$V$ loop can be on-shell, the imaginary component of the annihilation amplitude grows, as per the optical theorem. Second, when both the vector and fermion are present in the early universe, new annihilation channels are available, \eg semi-annihilation \cite{D'Eramo:2010ep}. We further expect that the fermion will often develop a nonnegligible contribution to the thermal relic, if light enough.

Ref.~\cite{DiFranzo:2015nli} showed how the SM gauge interactions of the fermions could play an important role in setting the relic abundance in this model. In that work, the gauge interactions presented themselves in box diagrams connecting external legs such as $V$-$V$-$Z$-$Z$ and $V$-$V$-$W$-$W$. When dark matter is heavy enough, these processes further increase the dark matter annihilation cross-section. However, in the present work we wish to focus on the role that the fermions could play in setting the relic abundance as dark matter itself or at least as a degree-of-freedom present in the early universe. In order to better isolate this phenomena from the SM gauge interactions, we will primarily be interested on the part of parameter space where the SM gauge interactions are subdominant to the Higgs interactions. Further, we will make the additional simplifying assumptions that the lightest fermion is the only relevant fermion for the phenomenology and that the scalar degree of freedom may be ignored. This is essentially the ``Single Fermion Limit'' explored in Sec.~III.A. of Ref.~\cite{DiFranzo:2015nli}. It is important to note that the above assumptions tend to be conservative, as including effects from the other fermions and their gauge interactions most often reduce the relic abundance with minimal changes to other observables, further opening up viable parameter space.

The SM gauge interactions will not be completely ignored. A coupling between the fermion and the $Z$ boson, can have marked effects. This coupling can be very small, in fact choosing $y_\chi=y_\psi$ will only generate off-diagonal couplings between the neutral fermions and $Z$ boson, without appreciably decreasing the corresponding Higgs couplings, \eg see the third set of benchmark parameters in Ref.~\cite{DiFranzo:2015nli}. This alignment may need to be highly tuned to avoid the relevance of the $Z$ boson, therefore we will investigate the phenomenological effect of this coupling. It is important to note that the diagonal coupling of the $Z$ to the fermions is only axial. This can be see from CC$^\prime$ symmetry. Taking $\Psi$ to be a neutral fermion, we find that $\bar{\Psi}\gamma_\mu \Psi$ and $V_\mu$ are odd under CC$^\prime$, whereas $\bar{\Psi}\gamma_\mu \gamma_5 \Psi$ and $Z_\mu$ are even. Therefore, the $Z$ can only have an axial coupling to a particular new fermion.

For the remainder of the paper, we will denote the vector field as $V$ and the lightest new fermion as $N_1$. The subscript on the fermion serves as a reminder that it is the lightest neutral state. Therefore we will be concerned
with five parameters in our study: 
\begin{itemize}
\item $M_V$: mass of vector, $V$
\item $M_{N_1}$: mass of fermion, $N_1$
\item $g_V$: $U(1)^\prime$ gauge coupling
\item $Y_N$: effective yukawa coupling of $N_1$ to the Higgs
\item $c_z$: parameter for $N_1$ coupling to $Z$ boson
\end{itemize}
The simplified interaction Lagrangian is given by:
\begin{equation}
\mathcal{L} \supset ~ g_V V^\mu \bar{N_1}\gamma_\mu N_1 ~+~ \frac{Y_N}{\sqrt{2}} h \bar{N_1}N_1 ~+~ \frac{e c_z}{2 c_w s_w} Z^\mu \bar{N_1}\gamma_\mu \gamma_5 N_1
\label{eq:simplag}
\end{equation}
Here the $Z$ coupling has been normalized such that $|c_z|\leq1$.

Where necessary, we utilize FeynArts \cite{Hahn:2000kx}, FormCalc, and LoopTools \cite{Hahn:1998yk} to ensure that the full momentum and mass dependence of the loop-level processes are properly taken into account. For vector annihilation, this includes the box diagrams which become relevant above the two Higgs final state threshold. The full loop dependence was incorporated into micrOMEGAS \cite{Belanger:2014vza} to correctly account for the temperature dependence of the annihilation cross-section.


\section{Thermal History of the Two Component System}
\label{sec:abundance}

\begin{figure}
$\vcenter{\hbox{\includegraphics[width=0.49\linewidth]{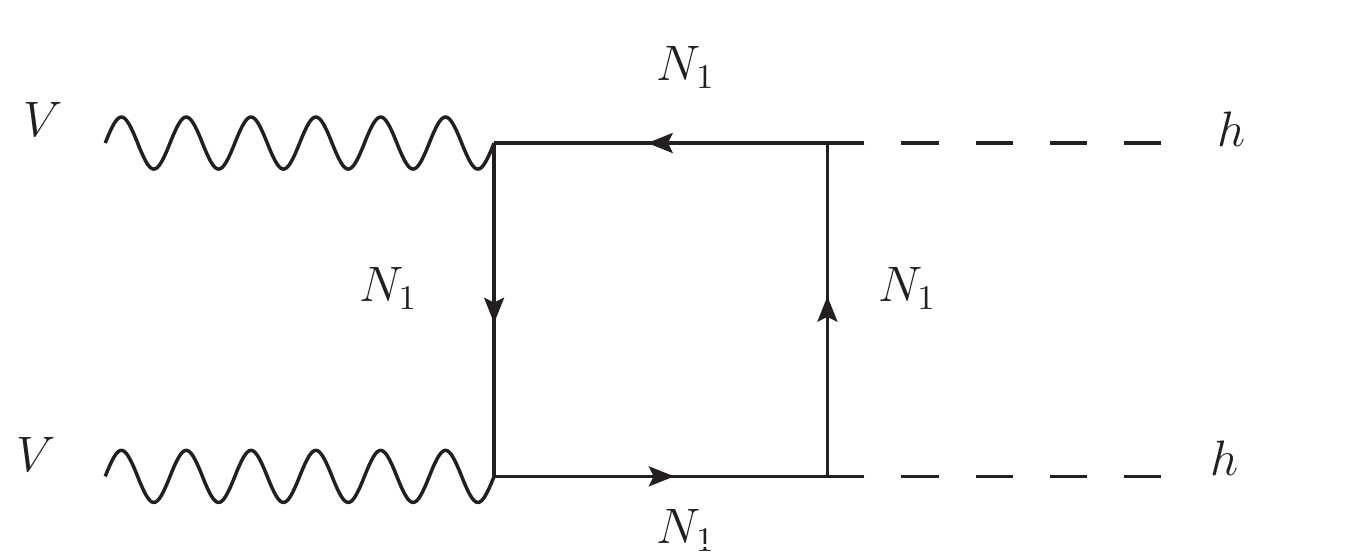}}}$
$\vcenter{\hbox{\includegraphics[width=0.49\linewidth]{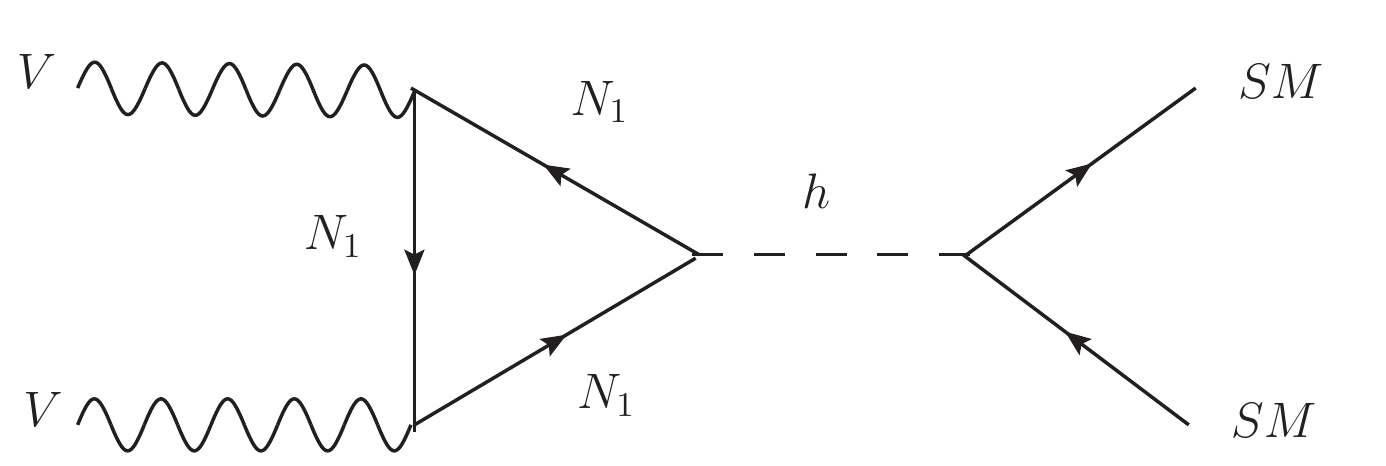}}}$

\includegraphics[width=0.38\linewidth]{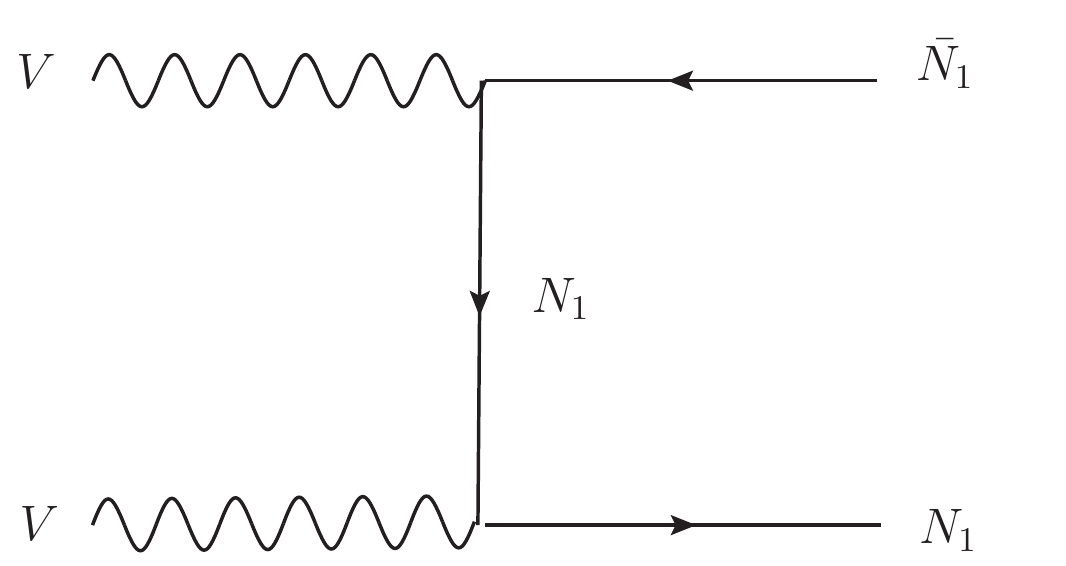}
\caption{Feynman diagrams showing the most dominant annihilation processes for the vector dark matter.}
\label{fig:annih0}
\end{figure}

\begin{figure}
\begin{center}
\includegraphics[width=0.45\linewidth]{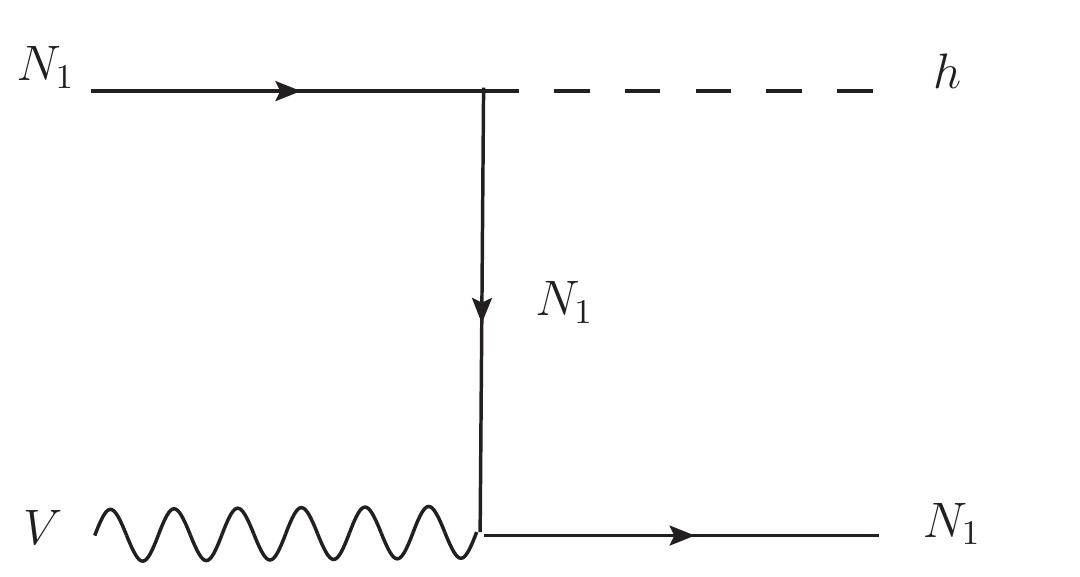}
\includegraphics[width=0.49\linewidth]{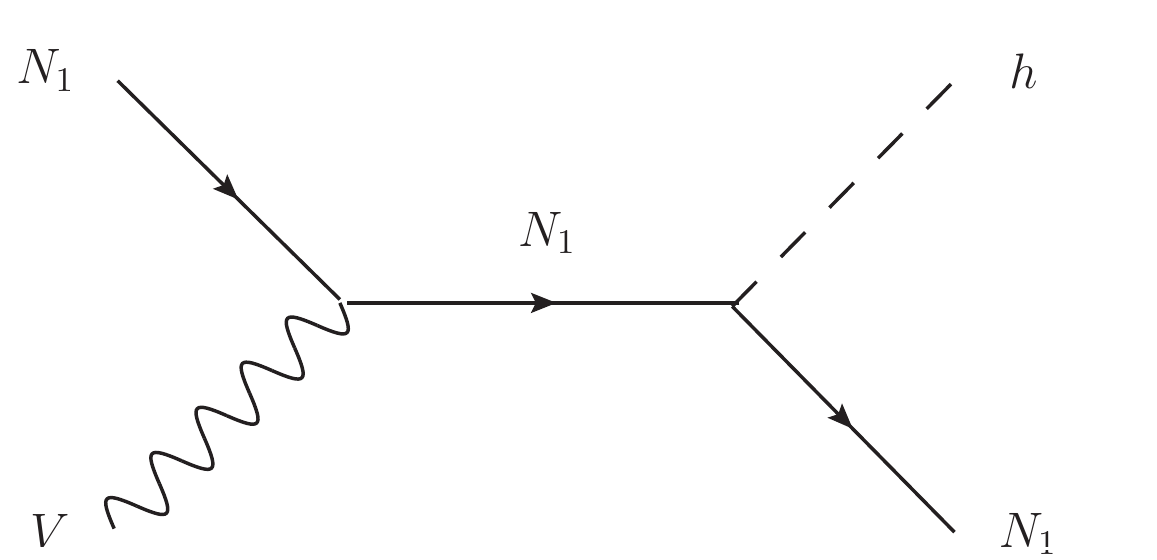}
\end{center}
\caption{Representative diagrams showing the semi-annihilation processes, relevant when the vector and the fermion have similar masses. Note that when considering couplings to the $Z$, similar diagrams exist with the $Z$ in place of the Higgs.}
\label{fig:annih1}
\end{figure}

The annihilation diagrams for the vector are given in Fig.~\ref{fig:annih0}. There are similar diagrams for the fermion; aside from cutting these loop diagrams, there is also a process through an s-channel $Z$ as well as $ZZ$ and $ZH$ final state channels. There are also semi-annihilation channels shown in Fig.~\ref{fig:annih1} and similar diagrams with the $Z$ in place of the Higgs.

There are three classes of interactions in our model. There are the usual annihilation channels where the final states are SM fields. There are processes that don't involve SM fields in the final state, which convert one species of dark matter into another. Finally, there are semi-annihilation processes where the final state has a DM particle and a SM particle.

This model has two distinct semi-annihilation channels. One reduces vector density without changing fermion density, $VN \rightarrow XN$. The second converts fermion density into vector, $N\bar{N}\rightarrow VX$. These rates will be most relevant when $X$ is on-shell, since $X$ must be the Higgs or $Z$, these rates are most relevant when $V$ and $N$ are relatively heavy. For $VN\rightarrow XN$, $M_V \gtrsim M_X$ whereas $N$ could be lighter, so long as the vector abundance is not too Boltzmann suppressed. For $N\bar{N}\rightarrow VX$, we find that $2M_{N_1} \gtrsim M_V+M_X$. Interestingly, this process can still be relevant for $V$--$N_1$ mass splittings which would normally suggest that co-annihilation is irrelevant. Specifically, if the vector is heavier than the fermion such that the vector abundance is highly Boltzmann suppressed, vectors may still be produced by this process thereby reducing the total abundance. This breaks the phenomenology into three distinct regimes, where ``much greater/less than" should be interpreted as one field's abundance being highly Boltzmann suppressed:

\begin{itemize}
\item \underline{$M_V \gg M_{N_1}$}: If $M_V$ is too large to significantly effect the freeze-out of the fermion, typically semi-annihilation is not relevant and conversion processes are not accessible. One caveat being processes such as $N_1 \bar{N}_1 \rightarrow V H/Z$, which can be relevant for mass differences larger than would be expected based on typical semi-annihilation processes. Eq.~\ref{eq:yabund} nearly reduces to that of a Fermion Higgs Portal. The vector relic abundance is increasingly small for larger $M_V$, however note that when $M_V>2M_{N_1}$, the vector is no longer stable and will not retain an abundance.

\item \underline{$M_V \ll M_{N_1}$}: Likewise, if the fermion is very heavy it will not significantly effect the current day relic abundance as a degree-of-freedom, again reducing to a single component DM scenario composed of vector DM. However, note that the fermion is still necessary for the vector's loop interaction with the SM. Therefore, this interaction will be suppressed for larger fermion masses, making it increasingly difficult for the vector to be a thermal relic.

\item \underline{$M_V \sim M_{N_1}$}: This scenario is the most phenomenologically rich. Here the masses are close enough that semi-annihilation and conversion processes may take place. The details of the freeze-out process will heavily depend on the couplings and masses chosen. It is this regime we wish to study in more detail in this work.
\end{itemize}

The evolution of the number density of dark matter is described by a set of coupled Boltzmann equations. These are parametrized in terms of the number of dark matter particles per comoving volume and entropy density of the Universe. The coupled Boltzmann equations for the different dark matter species is written as a function of the temperature $x = M_{N_1}/T$:
%
\begin{align}
x^2\frac{dY_{N_1}}{dx} &= - \lambda_{N_1 \bar{N_1} \rightarrow X X} \left[ Y_{N_1}^2 - (Y_{N_1}^{eq})^2 \right] - \lambda_{N_1 \bar{N_1} \rightarrow V V}  \left[Y_{N_1}^2 - \left(\frac{Y_{N_1}^{eq}}{Y_V^{eq}}\right)^2 Y_V^2 \right] \nl
& \qquad \qquad - \lambda_{N_1 \bar{N_1} \rightarrow V X} \left[ Y_{N_1}^2 - \frac{(Y_{N_1}^{eq})^2}{Y_V^{eq}} Y_V \right], \nl 
x^2 \frac{dY_V}{dx} &= - \lambda_{VV \rightarrow X X} \left[Y_V^{2} - (Y_V^{eq})^2 \right] - \lambda_{V V \rightarrow N_1 \bar{N_1}} \left[Y_V^2 -  \left(\frac{Y_{N_1}^{eq}}{Y_V^{eq}}\right)^2 Y_{N_1}^2 \right] \nl
- & \frac{1}{2}\lambda_{N_1 V \rightarrow N_1 X} ~ Y_{N_1} \left[Y_V - Y_V^{eq}\right] + \frac{1}{2}\lambda_{N_1 \bar{N_1} \rightarrow V X} \left[ Y_{N_1}^2 - \frac{(Y_{N_1}^{eq})^2}{Y_V^{eq}} Y_V \right].
\label{eq:yabund}
\end{align}
Where $\lambda_{ij \rightarrow kl} = \frac{s(x=1)}{H(x=1)} \langle \sigma v\rangle_{ij \rightarrow kl}$, with $\langle \sigma v\rangle_{ij \rightarrow kl}$ the thermally averaged annihilation cross-section of species $i$ and $j$ into species $k$ and $l$.
The quantity $s$ is the entropy density of the Universe, H is the Hubble parameter, and $Y_i^{eq}$ is the equilibrium number density per comoving volume for the different species:
\begin{align}
Y_V^{eq} = \frac{g_1}{g_{*s}} \frac{45}{4\pi^4} r^2 x^2 K_2 [r x],  ~~~Y_{N_1}^{eq} = \frac{g_2}{g_{*s}} \frac{45}{4\pi^4} x^2 K_2[x].
\end{align}
Here $g_1=3$ and $g_2=4$ are the number of internal degrees of freedom of the vector and fermion, respectively. $ r = M_V/M_{N_1} $ is the ratio of the masses and $K_2[x]$ is the modified Bessel function.
We obtain the solution for the coupled Boltzmann equations numerically, using the micrOMEGAS 4.2.5 package \cite{Belanger:2014vza}.

\begin{figure}
\begin{center}
\includegraphics[width=0.49\linewidth]{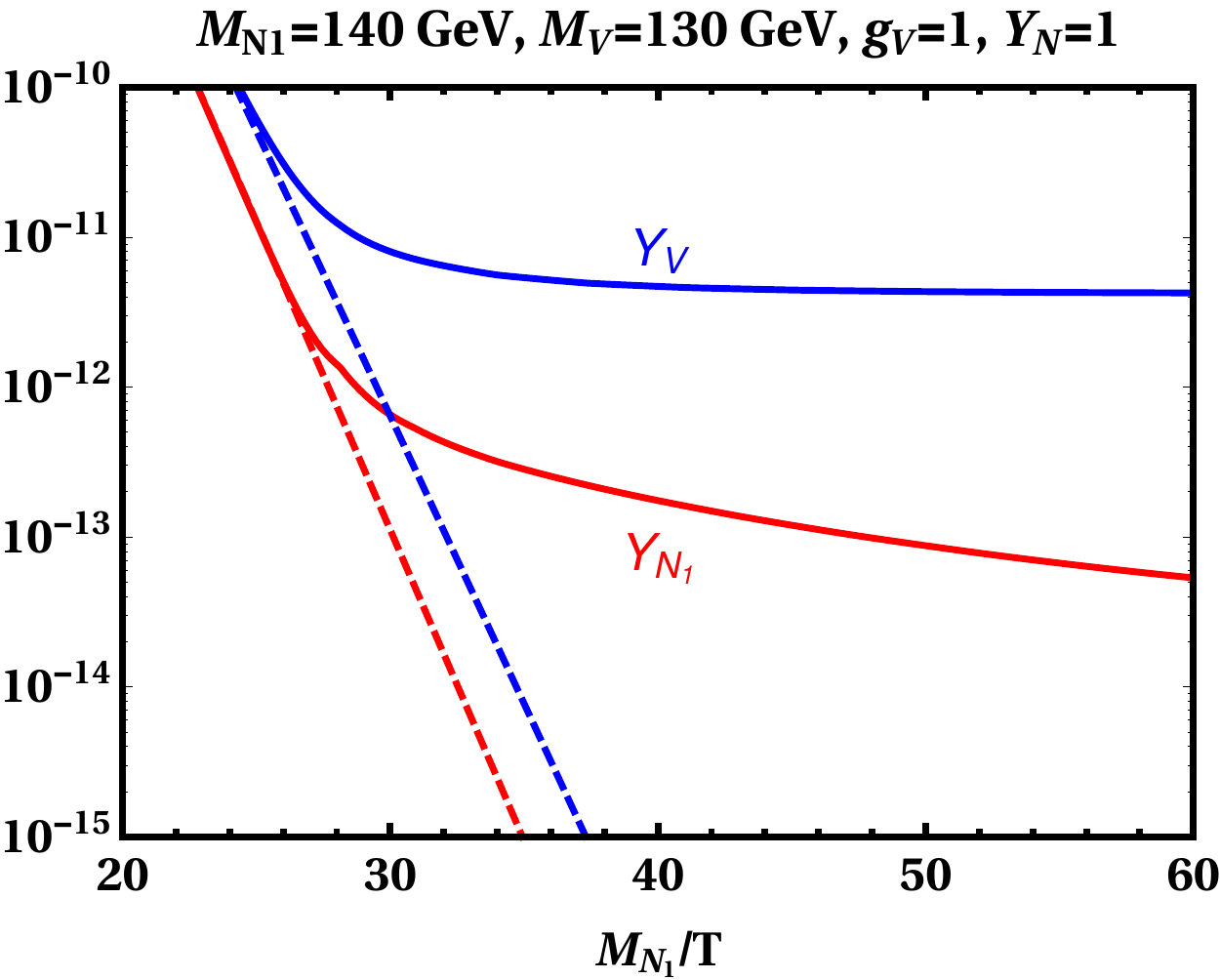}
\includegraphics[width=0.49\linewidth]{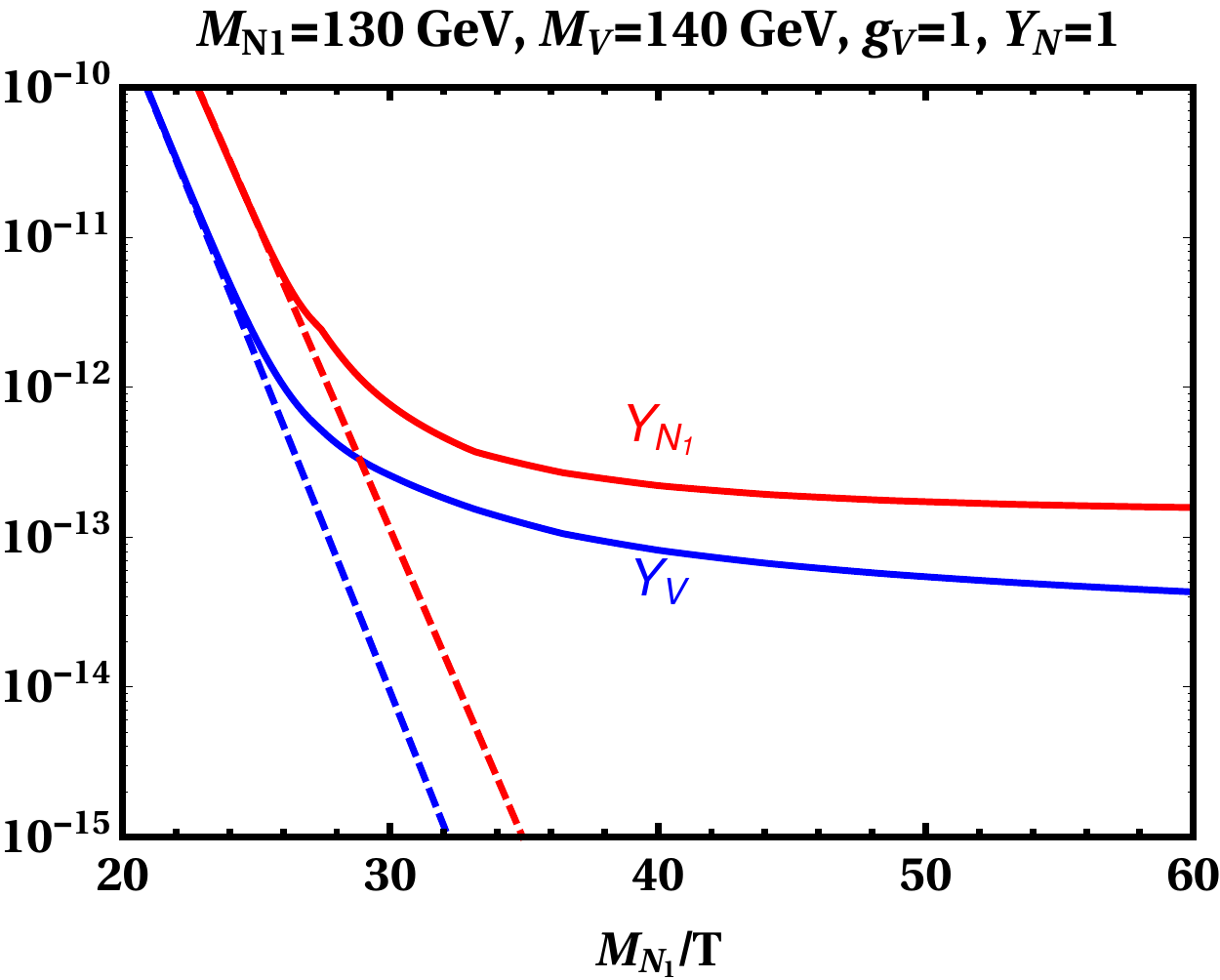}
\includegraphics[width=0.53\linewidth]{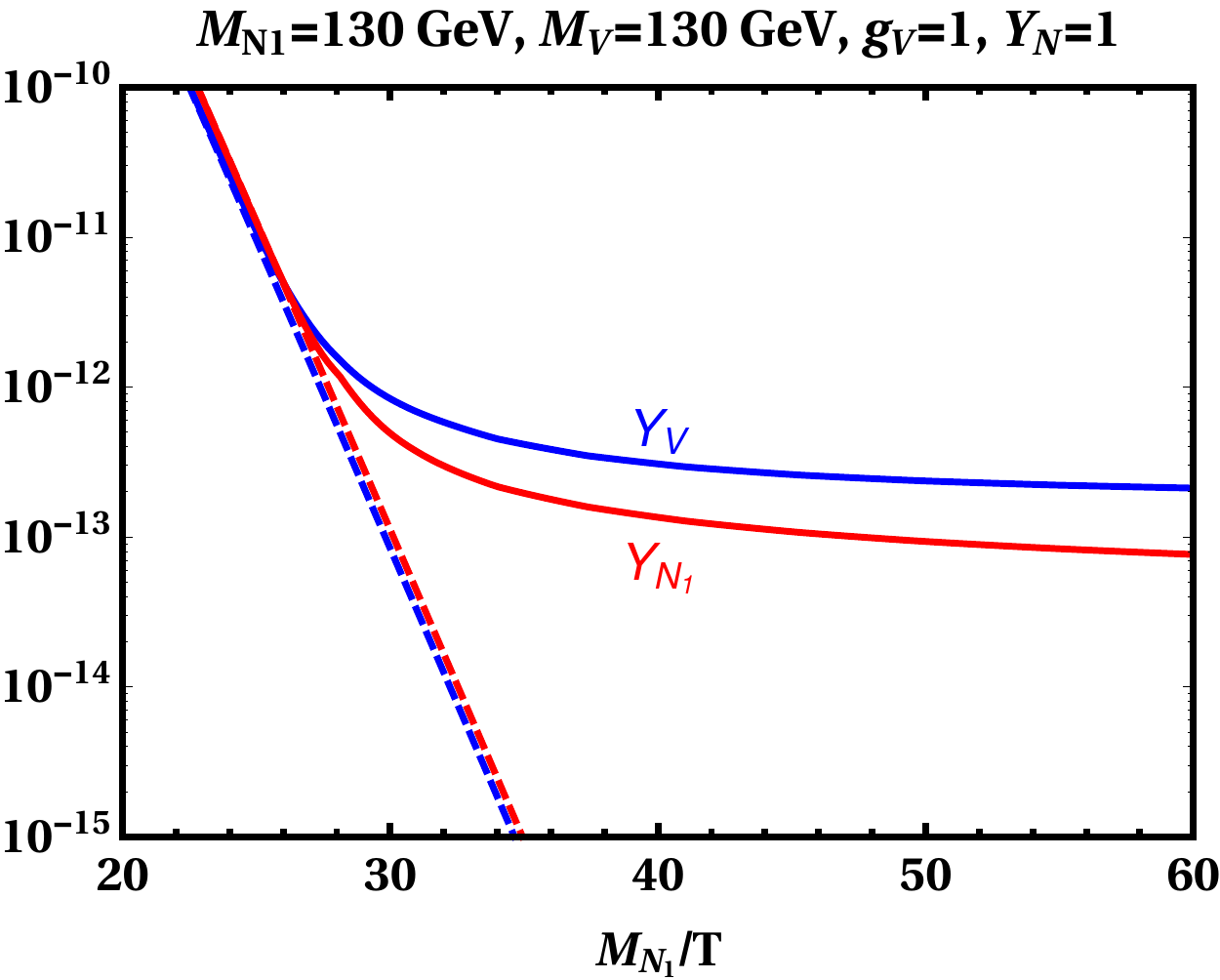}
\end{center}
\caption{Representative comoving number densities of the dark matter species after solving the coupled Boltzmann equations. The solid lines correspond to number densities and the dashed lines correspond to the equilibrium densities. The top two figures show the thermal histories when the lighter particle is the fermion and vice-versa. The bottom figure shows the thermal history when both the masses are degenerate. We choose $Y_N = 1$ and $g_V = 1$ as benchmark points.}
\label{fig:comoving}
\end{figure}

Typical thermal histories for the DM candidates are shown in Fig.~\ref{fig:comoving}. Note that even when the masses are degenerate, their respective thermal relics do not match. This is largely due to the fact that the vector couples to the SM at loop-level, therefore it annihilates at a slower rate and develops a greater thermal relic abundance. The presence of the fermion helps to maintain thermal equilibrium between the vector and the rest of the universe. However, upon freezing out, the fermion becomes a subdominant component of the total abundance. This phenomena is essentially the Assisted Freeze-out Mechanism \cite{Belanger:2011ww}.

\section{Phenomenology}
\label{sec:pheno}

Here we discuss the relevant sources of bounds, their corresponding formulae, and the methodology for setting limits.

\subsection{Relic Abundance}

\begin{figure}
\begin{center}
\includegraphics[width=0.494\linewidth]{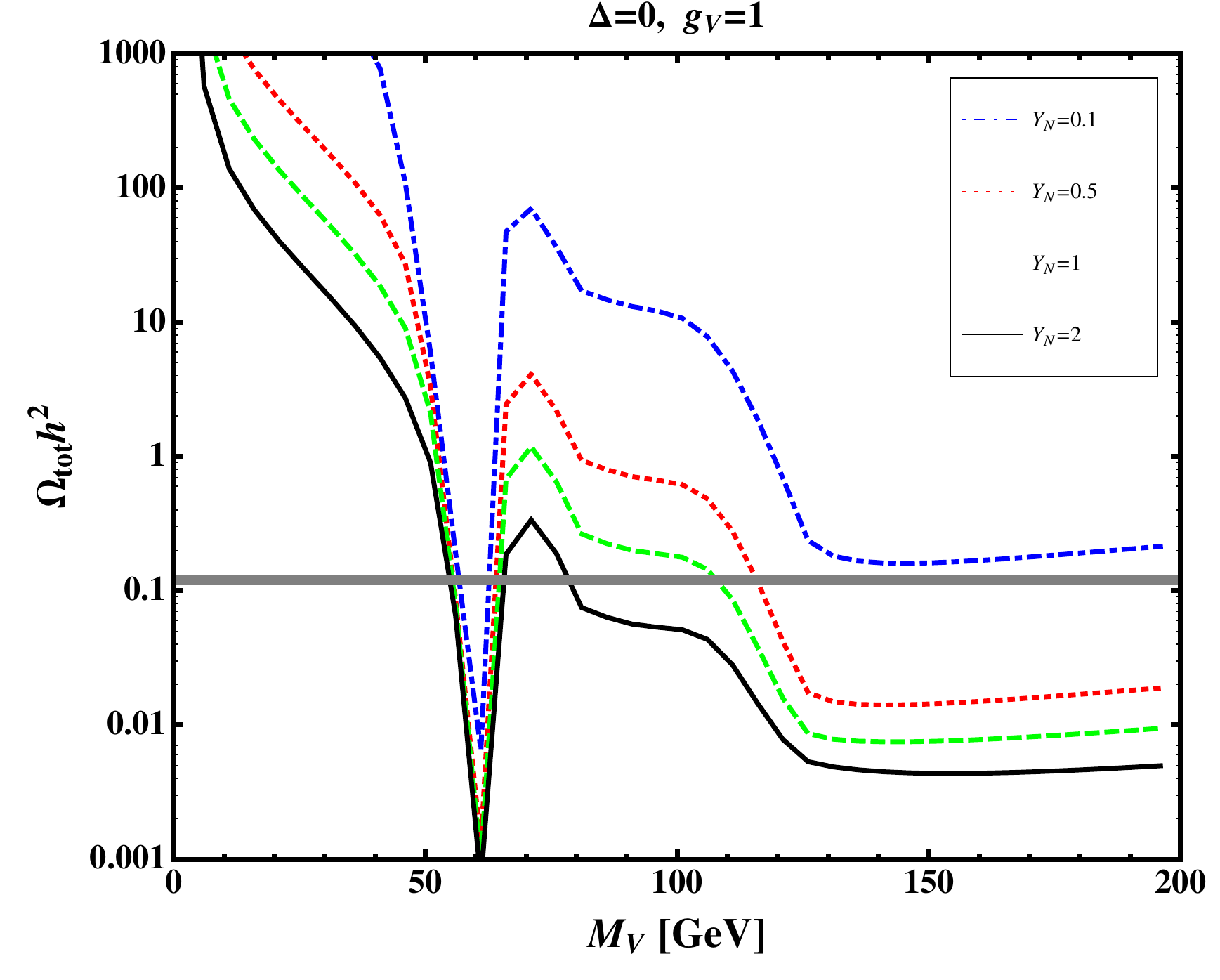}
\includegraphics[width=0.494\linewidth]{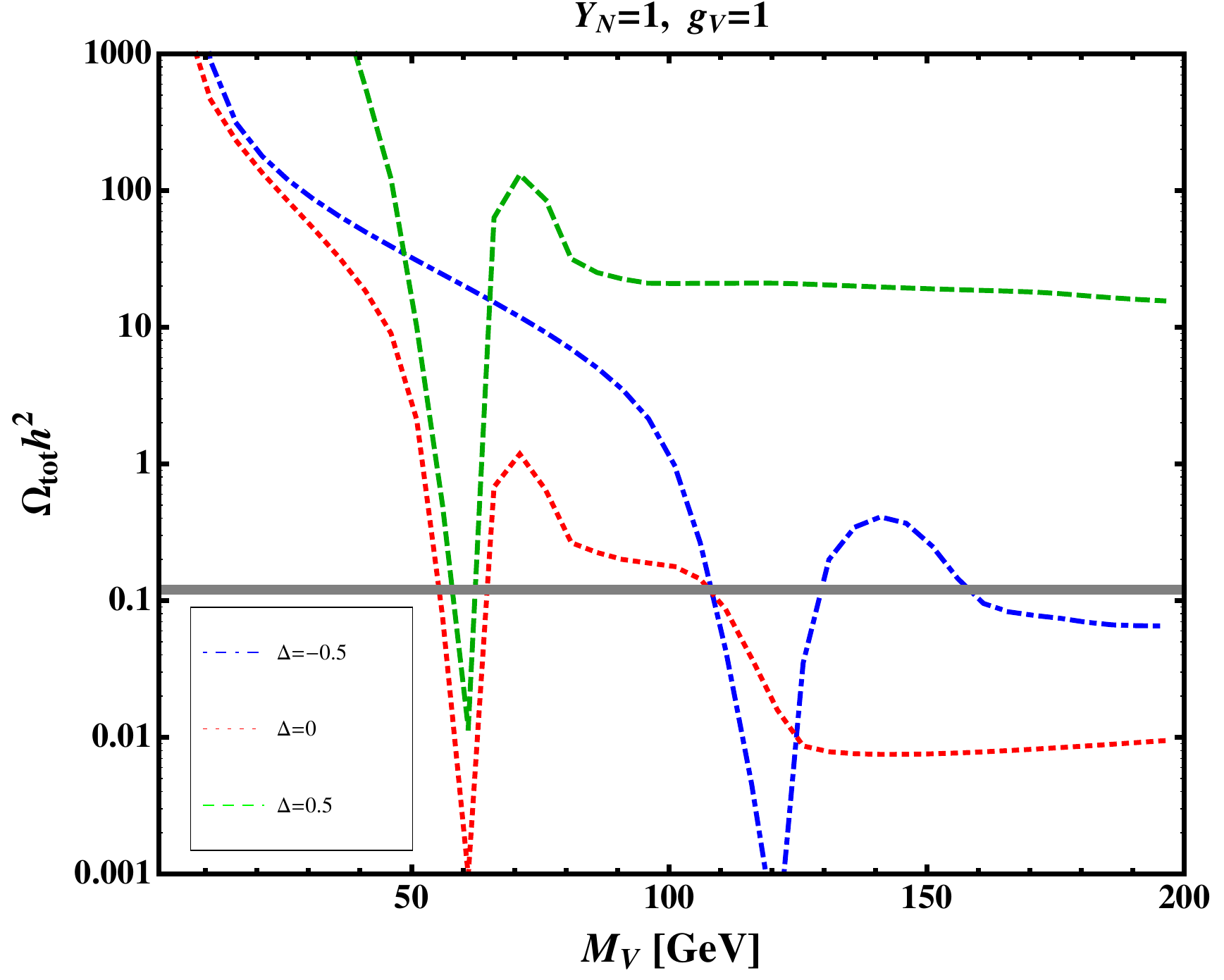}
\end{center}
\caption{Left: Relic density as function of the vector mass for different values of the Yukawa couplings and benchmark values for the mass splitting $\Delta$ and the gauge coupling $g_V$: The 
blue dashed curve represents $Y_N = 0.1$, red dotted $Y_N = 0.5$,  green dashed $Y_N = 1$ and orange solid $Y_N = 2$. Right: Relic density as a function of the vector mass for benchmark values of the 
Yukawa and gauge couplings for different values of the mass splitting $\Delta$; blue dot-dashed represents $\Delta = -0.5$, red dotted $\Delta = 0$ and green dashed $\Delta = 0.5$. The gray solid horizontal line represents 
the observed relic density of 0.12 \cite{Ade:2015xua}.}
\label{fig:relic_delta}
\end{figure}

\begin{figure}
\begin{center}
\includegraphics[width=0.6\linewidth]{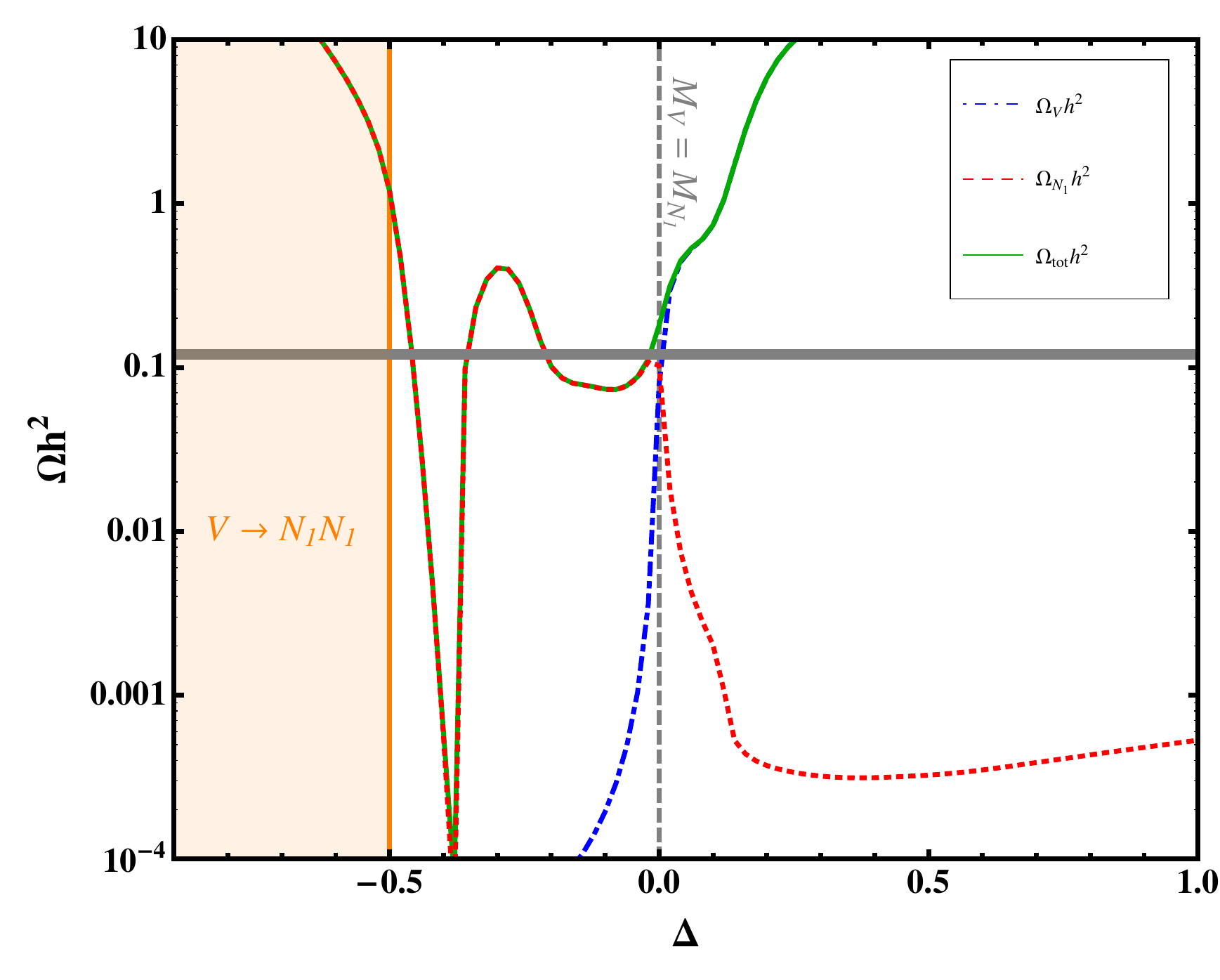}
\end{center}
\caption{The relative contributions of the vector and the fermion to the total relic density as a function of the mass splitting, for $g_V=1$, $Y_N=1$, and $M_V=100$ GeV. The red dotted curve represents the contribution of the Fermion, blue dot-dashed, the contribution of the vector and green solid is the total relic density of the two species. The orange shaded region shows where the vector is heavy enough to decay into the fermion, while the gray dashed vertical line shows the value of $\Delta$ for which the vector and the fermion are degenerate. The gray solid line, again represents the observed relic density.}
\label{fig:relic_delta_cont}
\end{figure}

In a scenario with multiple DM candidates, the relic density follows from the coupled Boltzmann equations, as discussed in Sec.~\ref{sec:abundance}, where the total predicted relic density from this model is the sum of the two components. However in order to examine the dependence of the relic abundance of each species on the parameters, 
we represent the relic abundance as a function of the masses of the DM states, as represented in Fig.~\ref{fig:relic_delta}. 
We define a mass splitting parameter:
\begin{equation}
\Delta = \frac{M_{N_1} - M_V}{M_V}.
\label{eq:splitting}
\end{equation}
Boltzmann suppression is determined by the relative mass difference, so $\Delta$ is useful as a crude measure of the relevance of co-annihilation processes. From Eq.~\ref{eq:splitting} we notice that for negative values of $\Delta$ the fermion is lighter and therefore is typically the dominant dark matter component. Furthermore, for $\Delta<-\nicefrac{1}{2}$, \ie the vector mass is more than twice the fermion mass, CC$^\prime$ no longer protects its stability and thus does not contribute to the total relic density. The transition in the relative contribution of each species to the relic abundance as a function of $\Delta$ is illustrated in Fig.~\ref{fig:relic_delta_cont}, where $M_V=100$ GeV.

Fig.~\ref{fig:relic_delta} shows the total relic density for various parameters. In the left panel, the large dip in the relic density is due to resonant annihilation through an s-channel Higgs into SM states, which is the dominant annihilation process in this mass regime. At slightly larger masses, the relic density decreases near the threshold for annihilating to $WW$ and $ZZ$. Another drop in the relic density occurs near the two Higgs final state threshold, which is mediated by both triangle and box diagrams represented in the top panels of Fig.~\ref{fig:annih0}. Semi-annihilation processes also become important in this high mass regime.

The right panel of Fig.~\ref{fig:relic_delta} shows the relic for various mass splittings. For negative $\Delta$, the fermion typically dominates, so the Higgs resonance will shift with the fermion mass accordingly, such that $M_{N_1} = M_h/2$. The positive $\Delta$ benchmark given, shows an absence of the $HH$ threshold. The large fermion mass running in the loop and the absence of a significant fermion relic density, suppresses processes of this form.

\subsection{Direct Detection}

\begin{figure}
\begin{center}
\includegraphics[width=0.39\linewidth]{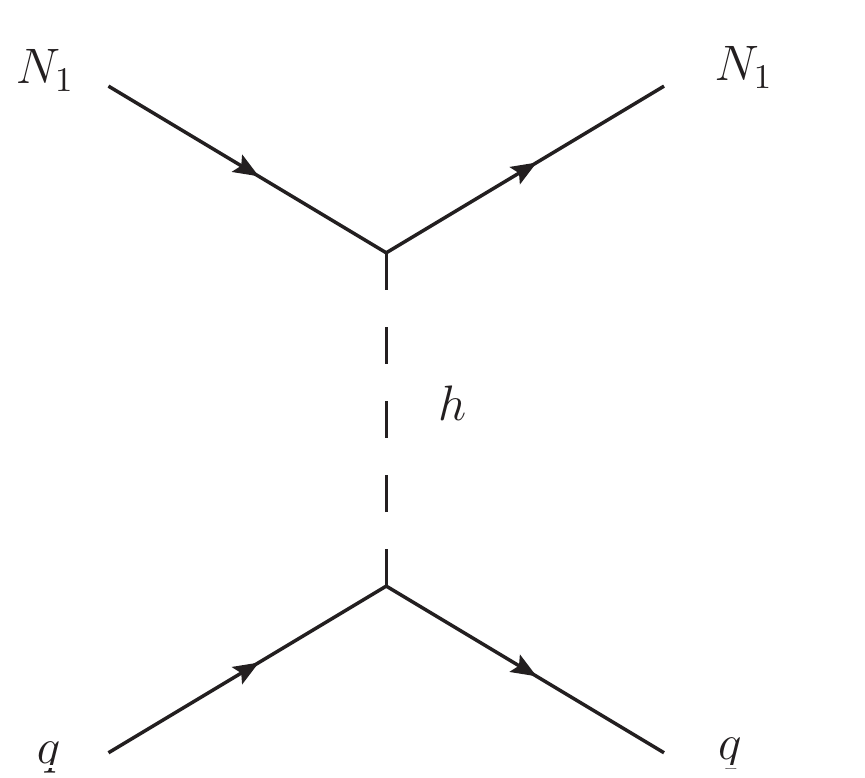}
\hspace*{0.1cm}
\includegraphics[width=0.45\linewidth]{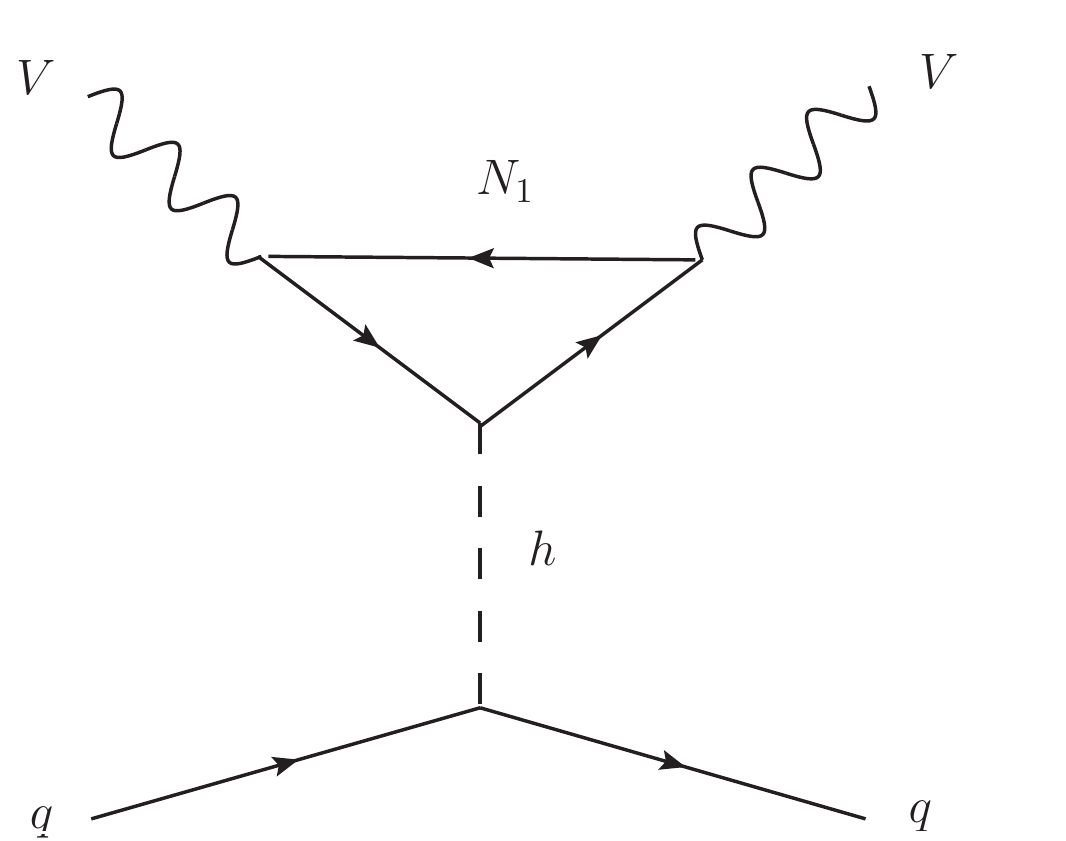}
\end{center}
\caption{Diagrams showing the scattering of the fermion and of the vector with SM quarks, in the left and right panels respectively. The scattering occurs through Higgs exchange and gives spin-independent rates.}
\label{fig:DD_scat}
\end{figure}

The vector and the fermion dark matter species interact with nucleons through Higgs exchange and thus the scattering cross-section is spin-independent. The scattering with nucleons is illustrated in Fig.~\ref{fig:DD_scat} and is calculated as:
\begin{equation}
\sigma_{SI}^V = \frac{Y_N^2 g_V^4 M_n^4}{4 \pi (M_n+M_V)^2 M_h^4} \frac{f_n^2}{v^2} |F_0(M_{N_1}, M_V)|^2,
\end{equation}
Where $F_0$ is a loop function defined in terms of the Passarino--Veltman coefficients and can be found in App.~\ref{app:appB}. The scattering of $N_1$ with nucleons occurs through tree-level Higgs exchange and is written as,
\begin{equation}
\sigma_{SI}^{N_1} = \frac{Y_N^2 M_{N_1}^2 M_n^4}{2 \pi (M_n + M_{N_1})^2 M_h^4} \frac{f_n^2}{v^2}
\end{equation}
$f_n$ are the nucleon matrix elements defined as
\begin{equation}
f_n = \sum_{q=u,d,s} f_{T_q}^{(n)} + \frac{2}{9} f_{T_G}^{(n)},
\end{equation}
We use the hadronic matrix elements $f_{T_{q}}$ obtained from DarkSUSY \cite{Gondolo:2004sc}. We define $v$ in the equations above as the standard model Higgs vacuum expectation value and $M_{n}$ the mass of the nucleon.

Current direct detection experiments have provided limits assuming that the local DM density consists of only one DM species. Thus, in a model with two DM candidates, those limits must be reinterpreted. In order to understand the limits set by experiments and properly apply them to our specific study, we consider the recoil rates measured by the direct detection experiments. The differential recoil rate on a target nucleus per recoil energy for a single DM particle scattering off a nucleus is defined as: 
\begin{align}
\frac{dR}{dE_R} = \frac{\sigma_{\chi N}^{(0)} \rho_{\chi}^{loc}}{2 M_{\chi} \mu_{\chi N}^2} F^2(E_R) I_{\chi}(E_R).
\label{eq:drdeone}
\end{align}
Where $E_R$ is the recoil energy of the target nucleus, $\sigma_{\chi N}^{(0)}$ is the DM--nucleus cross-section at zero momentum transfer, $\rho_{\chi}^{loc} = ~0.3$ GeV/cm$^3$ is the local energy density of dark matter. $\mu_{\chi N}$ is the reduced mass of the dark matter and Nucleus system, $F^{2}(E_{R})$ is the nuclear form factor which depends on the recoil energy $E_R$. $I_{\chi}(E_R)$ is the velocity integral assuming some velocity distribution of the galactic dark matter halo, this depends on the minimum velocity required for a DM particle to cause a recoil, $V_{min} = \sqrt{2E_R M_N/\mu_{\chi N}^2}$. 

The DM--Nucleus cross-section can be written in terms of the DM--nucleon scattering cross-section and atomic number $A$ as
\begin{align}
\sigma_{\chi N}^{(0)} = \frac{ \mu_{\chi N}^2}{\mu_{\chi n}^2} \sigma_{\chi n}^{SI}  A^2.
\label{eq:nucleus1}
\end{align}
Eq.~\ref{eq:drdeone} can thus be represented as a function of the DM--nucleon scattering cross-section $\sigma_{\chi n}^{SI}$.

On the other hand when considering multiple DM particles forming part of the DM halo in the Milky way galaxy, one has to take into account the nuclear scattering of each species in the detector, meaning we have to consider each particle's contribution to the local halo density and each particle's velocity distribution in the galactic halo. The total recoil rate then should account for each particle's recoil and thus is represented as:
\begin{align}
\frac{dR}{dE_R} = \sum_{i} \frac{\sigma_{i N}^{(0)} \rho_i^{loc}}{2 M_i \mu_{i N}^2} F^2(E_R) I_i(E_R).
\label{eq:drdemany}
\end{align}
Note that in general the local DM density need not have the same composition as the cosmological abundance. However, for simplicity we will assume that this is the case here, \ie $\rho_i^{loc}/\rho_{\chi}^{loc} \sim \Omega_i/\Omega_{\chi}^{tot}$, with $\Omega_{\chi}^{tot} h^2 = 0.12$.

Following the formalism of Dynamical Dark Matter in \cite{Dienes:2012cf} we obtain the recoil rates for our two component scenario as a function of the cross-section of each species scattering off nucleons. We represent the recoil rate, after taking into account the scattering from both species, as
\begin{align}
\frac{dR}{dE_R} = \frac{ \rho_{\chi}^{loc} A^2}{2} \bigg[ \frac{\Omega_Vh^2}{0.12}\frac{~\sigma_{V n}^{SI}}{ \mu_{V n}^2 M_V }  I_V(E_R) + \frac{\Omega_{N_1}h^2}{0.12} \frac{~\sigma_{N_1 n}^{SI}}{ \mu_{N_1 n}^2 M_{N_1} }  I_{N_1}(E_R)  \bigg] F^2(E_R).
\label{eq:drde2comp}
\end{align}
Here $\sigma_{V n}^{SI}$ and $\sigma_{N_1 n}^{SI}$ are the spin-independent DM--nucleon scattering cross-sections for the vector and the fermion species respectively, while $I_V$ and $I_{N_1}$ represent the velocity distributions of each of the species in the galaxy. From Eq.~\ref{eq:drde2comp}, we find that the two species have a nontrivial effect on the recoil spectra. To properly set direct detection limits on a two-species scenario, the full predicted recoil spectra should be compared to data. However, very often there is a large hierarchy in the scattering rates of the two species, such that one dominates the total scattering rate. If this is the case, an approximate limit may be set by requiring each species to independently satisfy:
\begin{align}
\sigma_{DD}^{SI} ~>~ \frac{\Omega_i h^2}{0.12} ~\sigma_{i n}^{SI}.
\label{eq:sigluxcomp}
\end{align}
Here $\sigma_{DD}^{SI}$ is the limit on the DM--nucleon scattering cross-section quoted by a direct detection experiment, such as LUX. $\sigma_{i n}^{SI}$ is the scattering cross-section between species $i$ and a nucleon, predicted by the model. In the above, the predicted scattering cross-section has been weighted by the fractional abundance of that species in accordance with Eq.~\ref{eq:drde2comp}.

Eq.~\ref{eq:sigluxcomp} breaks down if the scattering rates of each species are similar, which would lead to a limit that is conservative by a factor of two, at most. Even though the scattering cross-sections are always hierarchical due to the vector scattering at loop level, the rates may still be similar if the vector has a relative abundance which compensates this hierarchy. However, it is a rare occurrence for the rates to be similar and near the edge of being excluded by this method. Since the relic abundance changes rapidly with the two masses, the direct detection exclusion curves would only shift by a small degree for these cases. For the successful benchmark parameters presented in the paper, this shift does not encroach into regions which would otherwise not be excluded.

There may very well be multiple contributions to the total dark matter relic abundance, of which this model may only explain a fraction. Therefore, we do not require the sum of the two species to compose the entirety of the dark matter relic abundance and only require that it not exceed the measured value. We use the most recent direct detection limits set by LUX \cite{Akerib:2016vxi} \footnote{We point out here that the limits from the experiments are not rescaled, since these are what the experiments report assuming one DM component.}.

\subsection{Invisible Higgs Width}
If the mass of either of the DM species is lighter than half the Higgs, i.e. $M_i < M_h/2$, then that species will contribute to the Higgs width. The Higgs partial width into vectors is given by:
\begin{align}
\Gamma_{h\rightarrow V V} = \frac{Y_N^2 g_V^4 \sqrt{1 - 4 M_V^2/M_h^2}}{64 \pi M_h} \Bigg[ M_h^4 |A_{inv}|^2 \bigg( 1 - 4 \frac{M_V^2}{M_h^2} + 6 \frac{M_V^4}{M_h^4} \bigg) 
~~~~~~~~~~~~~~~~~~~~~~ \nl
+ ~6 ~{\rm Re}[A_{inv}^* B_{inv}] ~M_h^2 ~\bigg(1 - \frac{2 M_V^2}{M_h^2} \bigg) + \frac{1}{2} |B_{inv}|^2  \frac{M_h^4}{M_V^4} \bigg(1 - 4 \frac{M_V^2}{M_h^2} + 12 \frac{M_V^4}{M_h^4} \bigg) \Bigg].
\end{align}
Where $A_{inv}$ and $B_{inv}$ are functions of the vector, fermion, and Higgs masses, the functional form of which can be found in App.~\ref{app:appA}.

The decay channels of the Higgs are further opened as it can also decay into the fermion, $N_1$, with the decay width:
\be
\Gamma_{h \rightarrow N_1 N_1} = \frac{Y_N^2 M_h}{16 \pi} \bigg(1 - 4~ \frac{M_{N_1}^2}{M_h^2} \bigg)^{3/2}.
\ee
Thus the total contribution to the invisible Higgs width becomes, 
\be
\Gamma_{h \rightarrow inv} = \Gamma_{h\rightarrow V V} + \Gamma_{h \rightarrow N_1 N_1}.
\ee
The ATLAS collaboration constrains ${\rm Br}(h\rightarrow inv) < 0.23$ at $95\%$ CL with 4.7$fb^{-1}$ of data at 7 TeV and 20.3$fb^{-1}$ at 8 TeV \cite{Aad:2015pla}, which we use to constrain our parameter space.

\subsection{Z couplings}

Thus far the discussion has ignored couplings to the $Z$. For direct detection, the $Z$ only induces SD and velocity suppressed SI direct detection cross-sections due to its axial coupling to the fermion. Unless the lightest fermion has an exceptionally small yukawa coupling and large $c_z$, direct detection constraints are dominated by Higgs-exchange.

One may also consider constraints from the invisible $Z$ width when the fermion is kinematically accessible, where new contributions should not exceed 2 MeV \cite{ALEPH:2005ab,Agashe:2014kda}. The most stringent constraint on the coupling is in the limit where the fermion is massless, where the invisible $Z$ width requires $|c_z|\lesssim 0.08$.

The most significant impact will be on setting the relic abundance. Note that s-channel annihilation through the $Z$ to SM fermions is helicity suppressed, therefore the cross-section is suppressed by $m_f^2$ \cite{Basirnia:2016szw}. However, the s-channel annihilation through the Higgs has a similar suppression due to the SM yukawas. Therefore, such processes may be important even when the top quark is not kinematically accessible.

\section{Results and Discussion}
\label{sec:results}


\begin{figure}
\begin{center}
\includegraphics[width=0.45\linewidth]{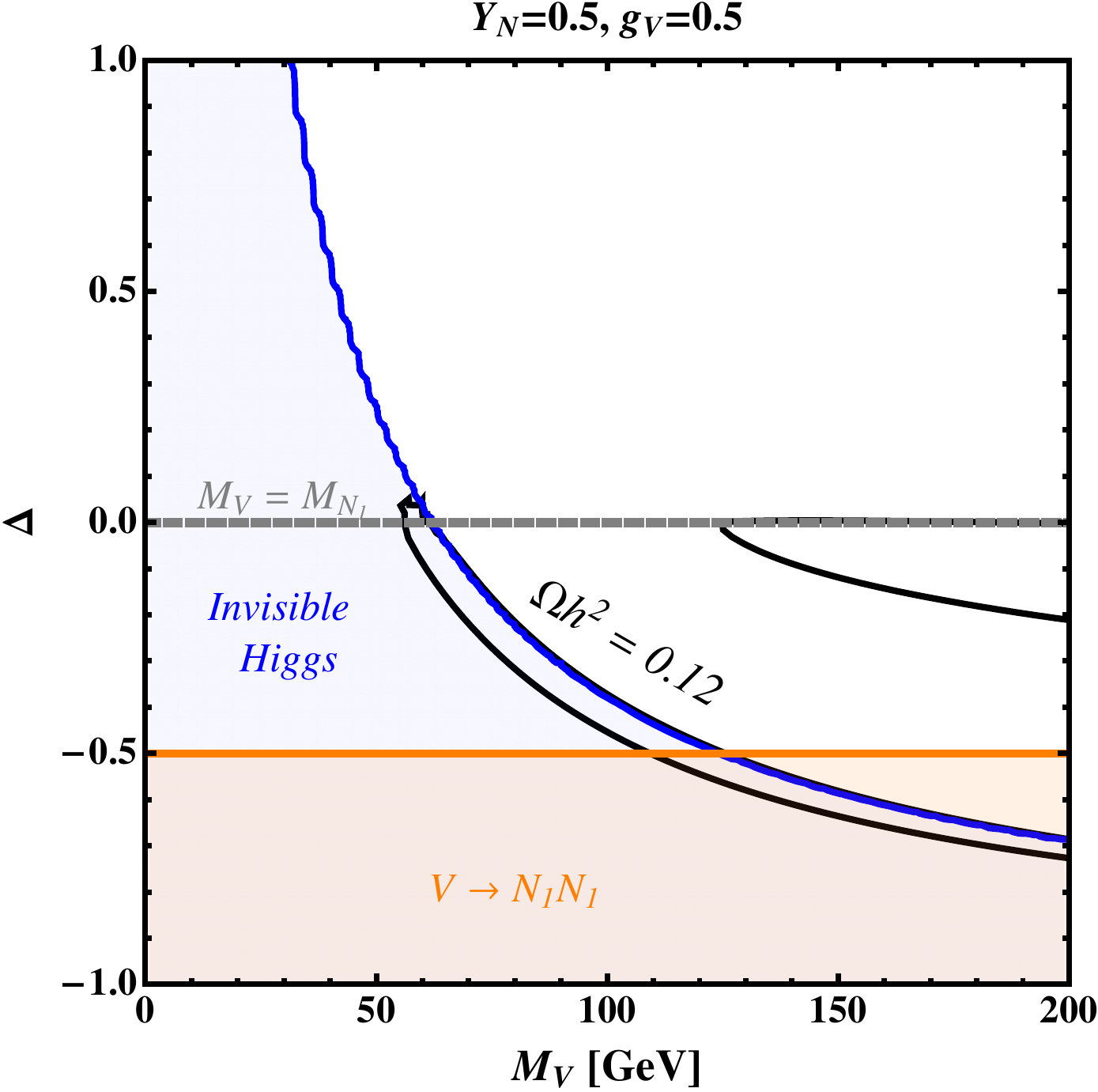}
\includegraphics[width=0.45\linewidth]{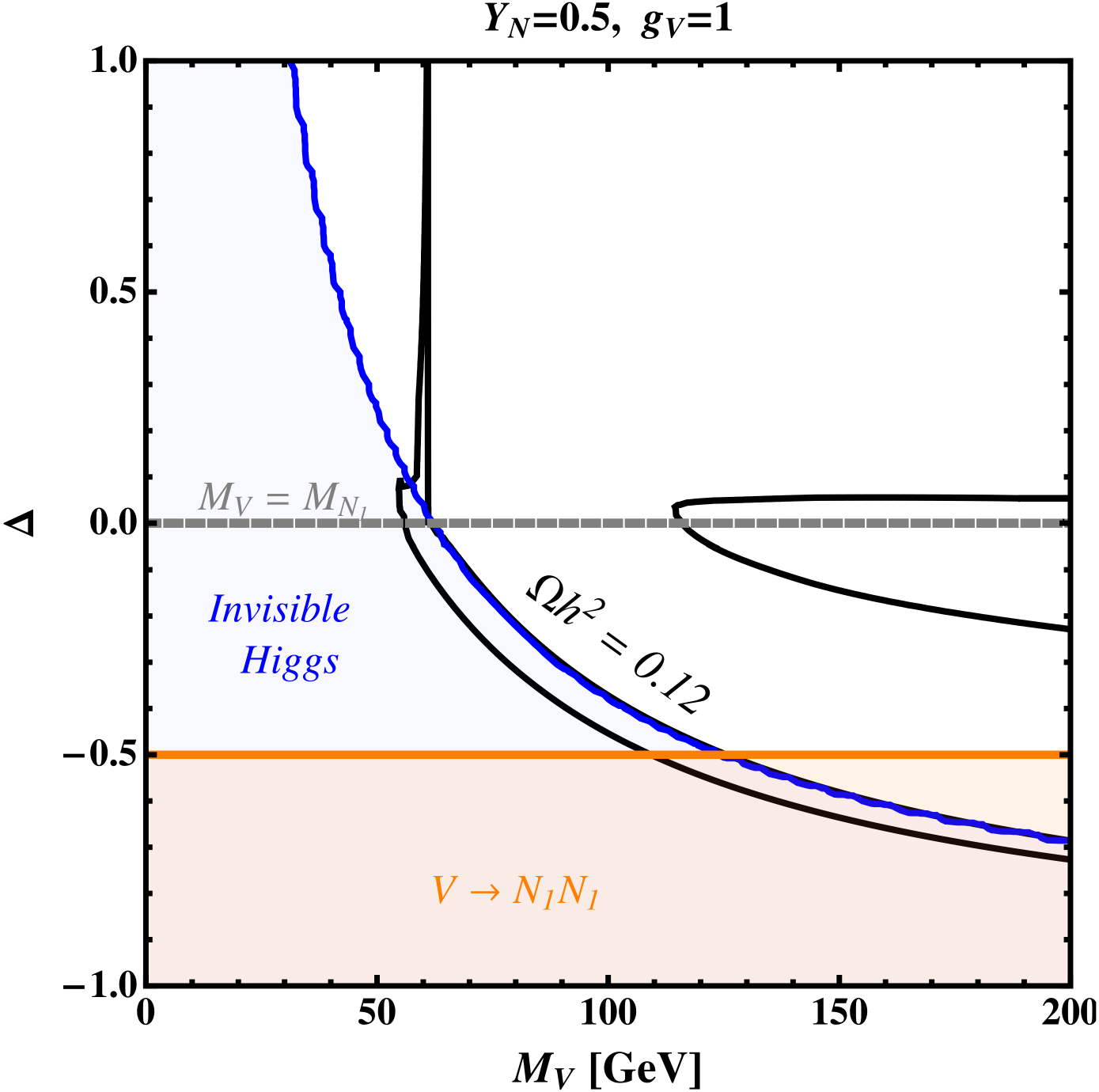}\\
\vspace{1.2ex}
\includegraphics[width=0.45\linewidth]{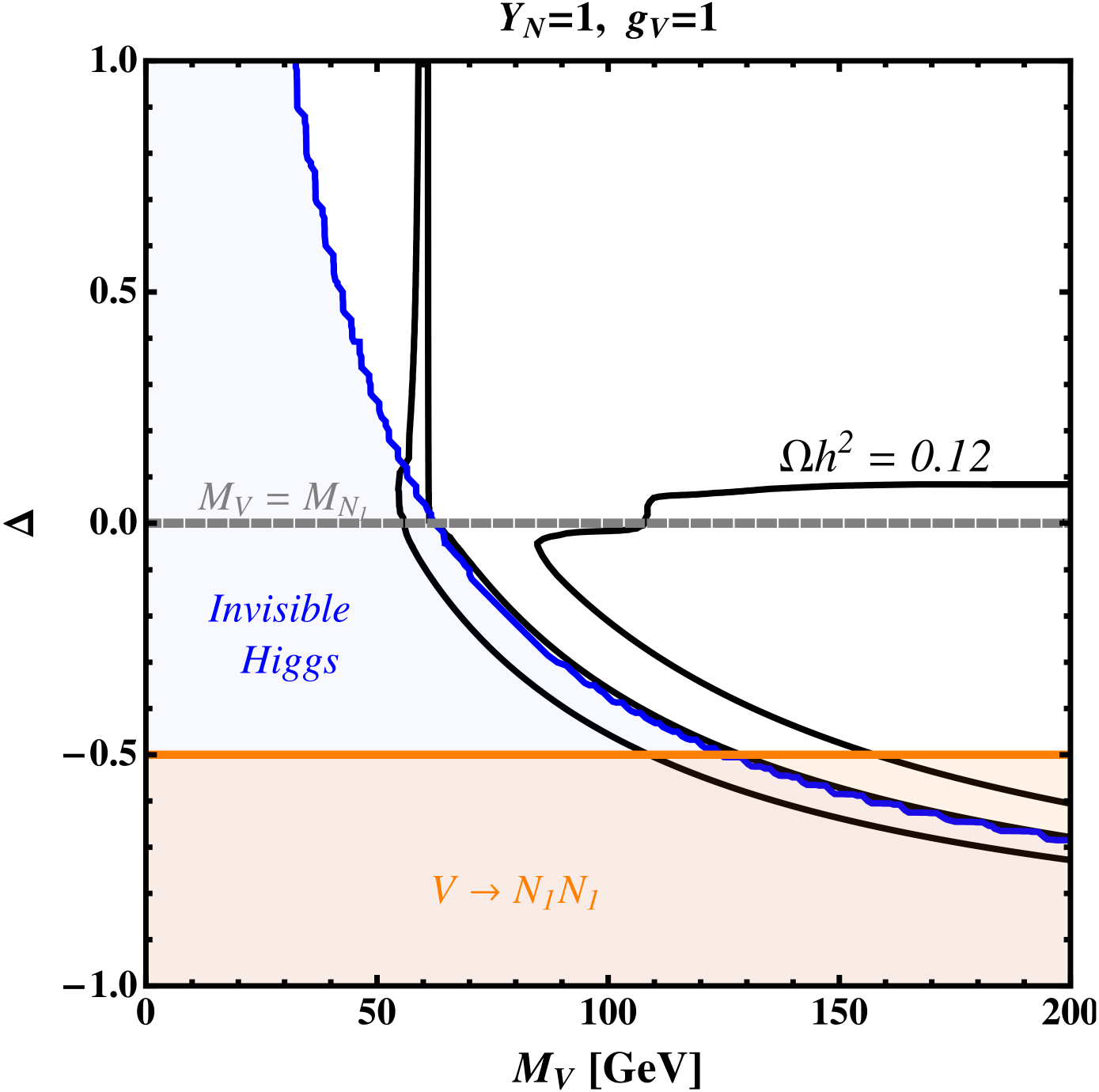}
\includegraphics[width=0.45\linewidth]{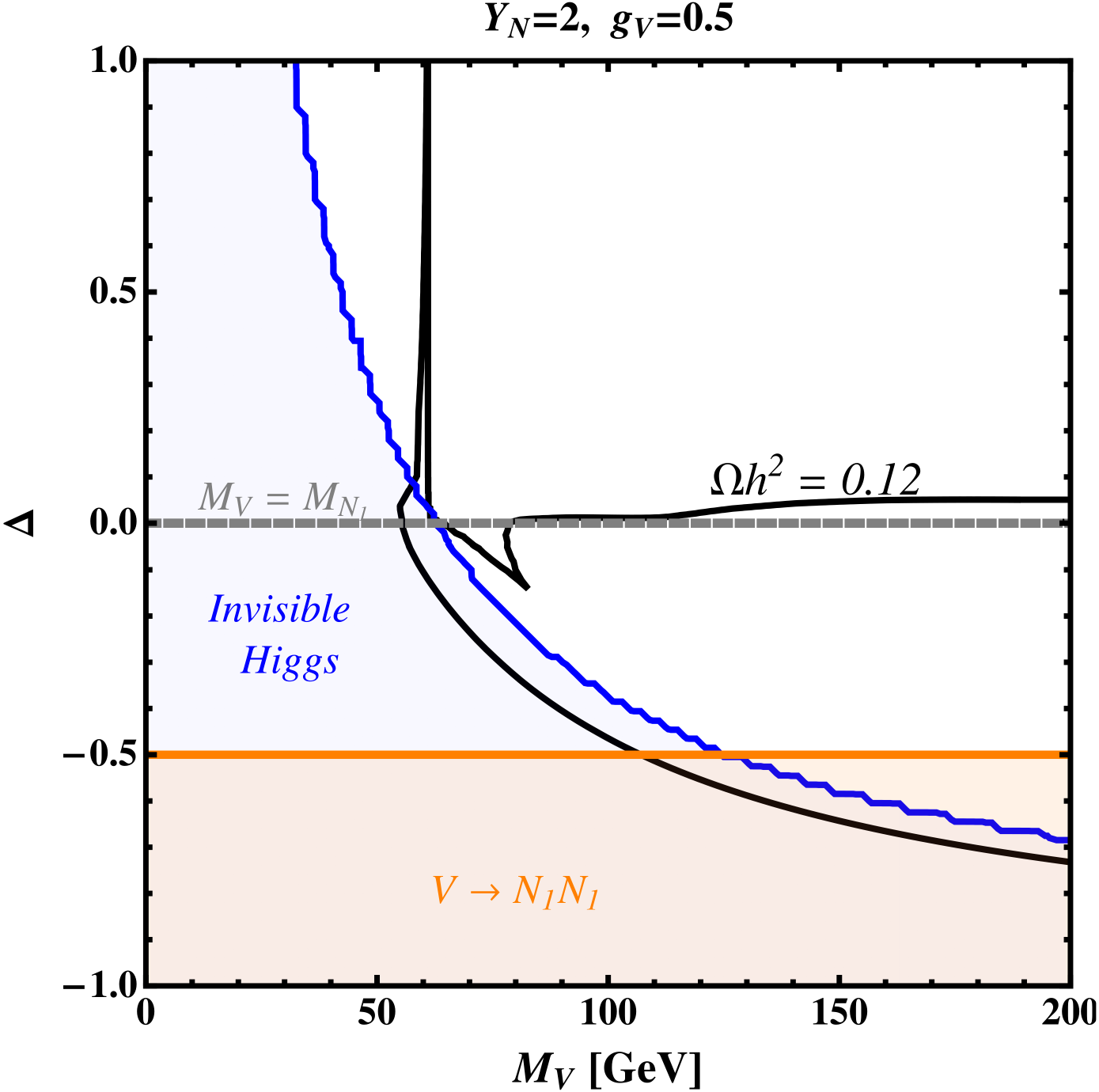}
\end{center}
\caption{Constraints from both relic abundance and Invisible Higgs measurements from the LHC, assuming $c_z=0$. The orange shaded region shows where the vector decays into the fermion and we effectively have only one thermal DM component contributing to the relic abundance. Along the solid black curves, we have a relic abundance in agreement with the observed cosmological dark matter density. The gray dashed line represents $\Delta = 0$ and roughly indicates the point where there is a transition of the relative contributions of each species to the thermal relic abundance. The blue shaded region indicates the limits from the invisible Higgs searches.}
\label{fig:limits}
\end{figure}

\begin{figure}
\begin{center}
\includegraphics[width=0.494\linewidth]{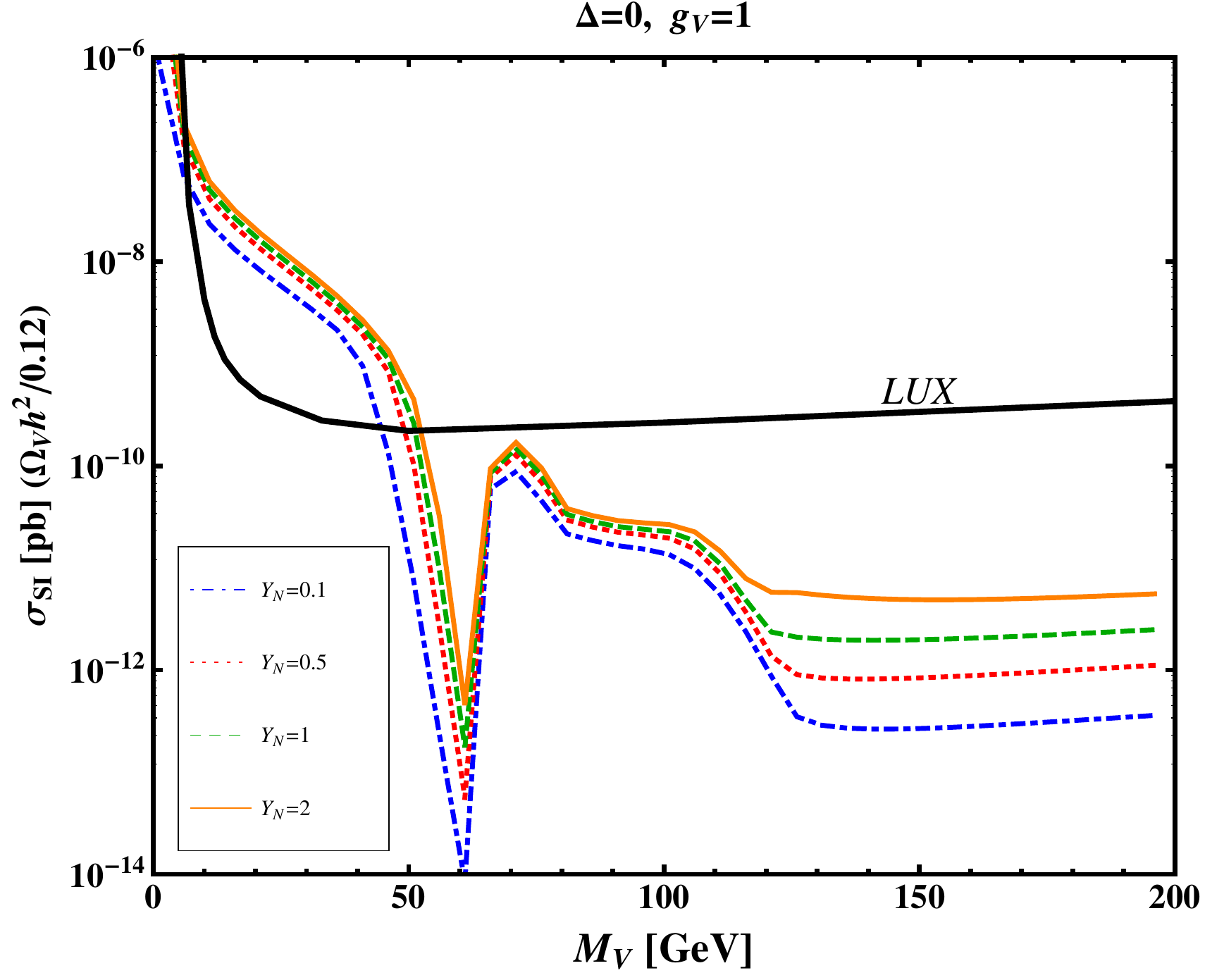}
\includegraphics[width=0.494\linewidth]{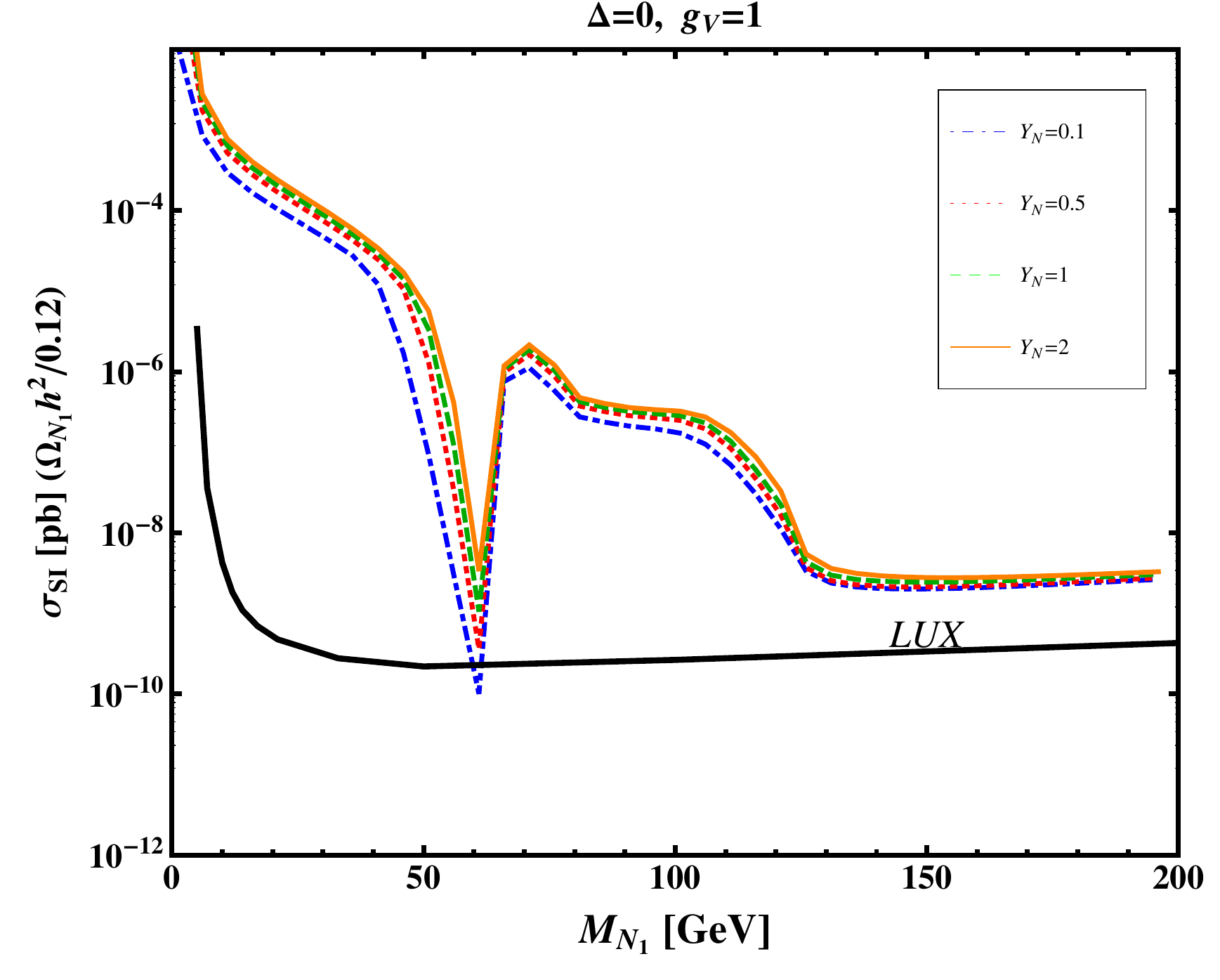}
\end{center}
\caption{Rescaled scattering cross-section (in pb) as a function of the mass of each DM species, (left) vector and (right) fermion. Each cross-section is rescaled according to its relative contribution to the observed relic density as indicated in Eq.~\ref{eq:sigluxcomp}. Each color line represents a different Yukawa coupling to the Higgs; for $Y_N = 0.1$ (blue dot-dashed), $Y_N = 0.5$ (red dotted), $Y_N = 1$ (green dashed) and 
$Y_N = 2$ (orange solid). The black solid line represents the most recent limits reported by LUX.}
\label{fig:DD_comp}
\end{figure}

In Fig.~\ref{fig:limits}, the contour of $\Omega h^2=0.12$ is shown in a solid black line for various benchmark parameters, as well as limits from the invisible Higgs width shown in blue. Regions which avoid these constraints lie inside the black contour and outside the blue shaded region.

The Higgs width excludes a large part of parameter space, nearly everywhere where the fermion is kinematically accessible by Higgs decay. For relic abundance, we find a thin curved region which corresponds to resonant fermion annihilation through the Higgs, which is almost entirely excluded by the Higgs width. The thin vertical region with positive $\Delta$ corresponds to resonant vector annihilation through the Higgs. There is also a region at larger vector mass where di-boson final states as well as semi-annihilation processes are kinematically accessible and dominate. Note that this region is mostly in the negative $\Delta$ regime where the vector abundance is Boltzmann suppressed due to its larger relative mass and rapid annihilation into fermions. The tree-level annihilation of the fermion allows for efficient annihilation. However, for the region near $\Delta=0$, semi-annihilation becomes important and even allows a thermal relic up to roughly $\Delta=0.1$ for some parameters.

Fig.~\ref{fig:DD_comp} shows the limits from direct detection for $\Delta=0$. Despite the fermion relic being a fraction of the total dark matter abundance, it is not enough to suppress direct detection constraints. For typical s-wave processes, reducing the coupling has a small effect on the predicted $\left<\sigma_{DD}\right> \times \Omega_{N_1} h^2$ since both the direct detection and annihilation cross-section scale with the coupling in the same way. However, the processes here are p-wave, where the relic abundance depends on a lower power of the freeze-out temperature compared to s-wave processes, \ie $\propto T_f^{-2}$ rather than $T_f^{-1}$. Therefore, variations in the freeze-out temperature are more apparent, leading to deviations from the approximation that the relic density is inversely proportional to the annihilation cross-section. In fact, decreasing the yukawa decreases the direct detection cross-section faster than it increases the relic abundance. In Fig.~\ref{fig:DD_comp}, we find that decreasing the yukawa can satisfy direct detection constraints. Unfortunately, in doing so this also causes these dark matter candidates to be overabundant. In fact there is very little room to simultaneously satisfy direct detection and be a thermal relic by solely altering the yukawa. The only region which typically evades bounds is positive $\Delta$ and $M_V \sim M_h/2$, due to the resonant annihilation of vectors through the Higgs.

This difficulty is largely due to the presence of the fermion, which has a large direct detection scattering cross-section. For fermion masses above 20 GeV, the fermion must satisfy $Y_N^2~\Omega h^2 \lesssim 10^{-4}$ in order to evade direct detection constraints. Masses below this are less constrained due to the direct detection threshold, however are heavily constrained by Higgs invisible constraints. Therefore, the fermion should be a subdominant component of the total dark matter abundance and/or have a small yukawa coupling.

Phenomenologically interesting regions, \ie where this model may explain a sizable portion of the dark matter abundance, are then restricted to scenarios where the dark matter is predominantly composed of the vector which has a comparably small direct detection cross-section.

\begin{figure}
\begin{center}
\includegraphics[width=0.49\linewidth]{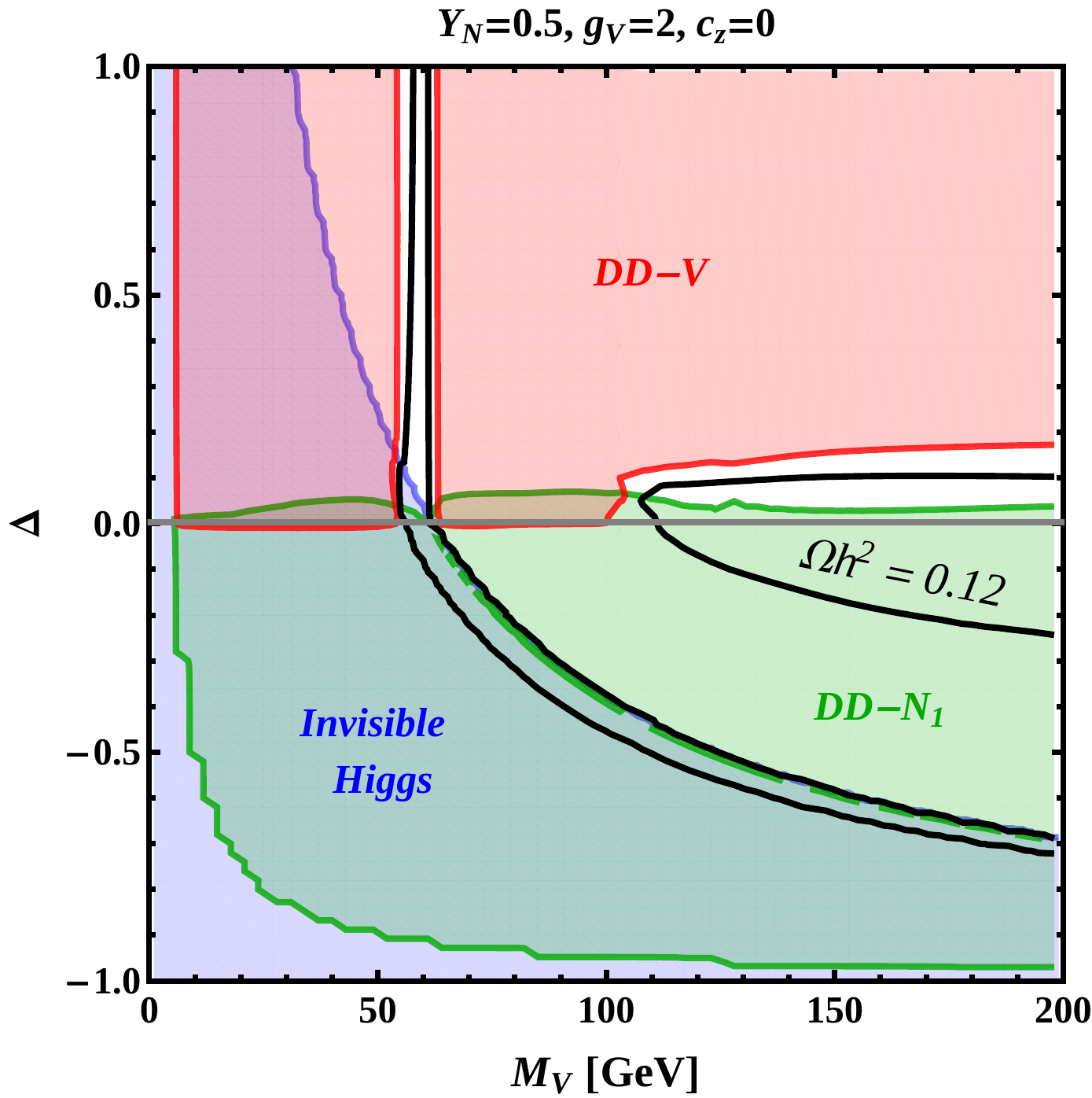}
\includegraphics[width=0.49\linewidth]{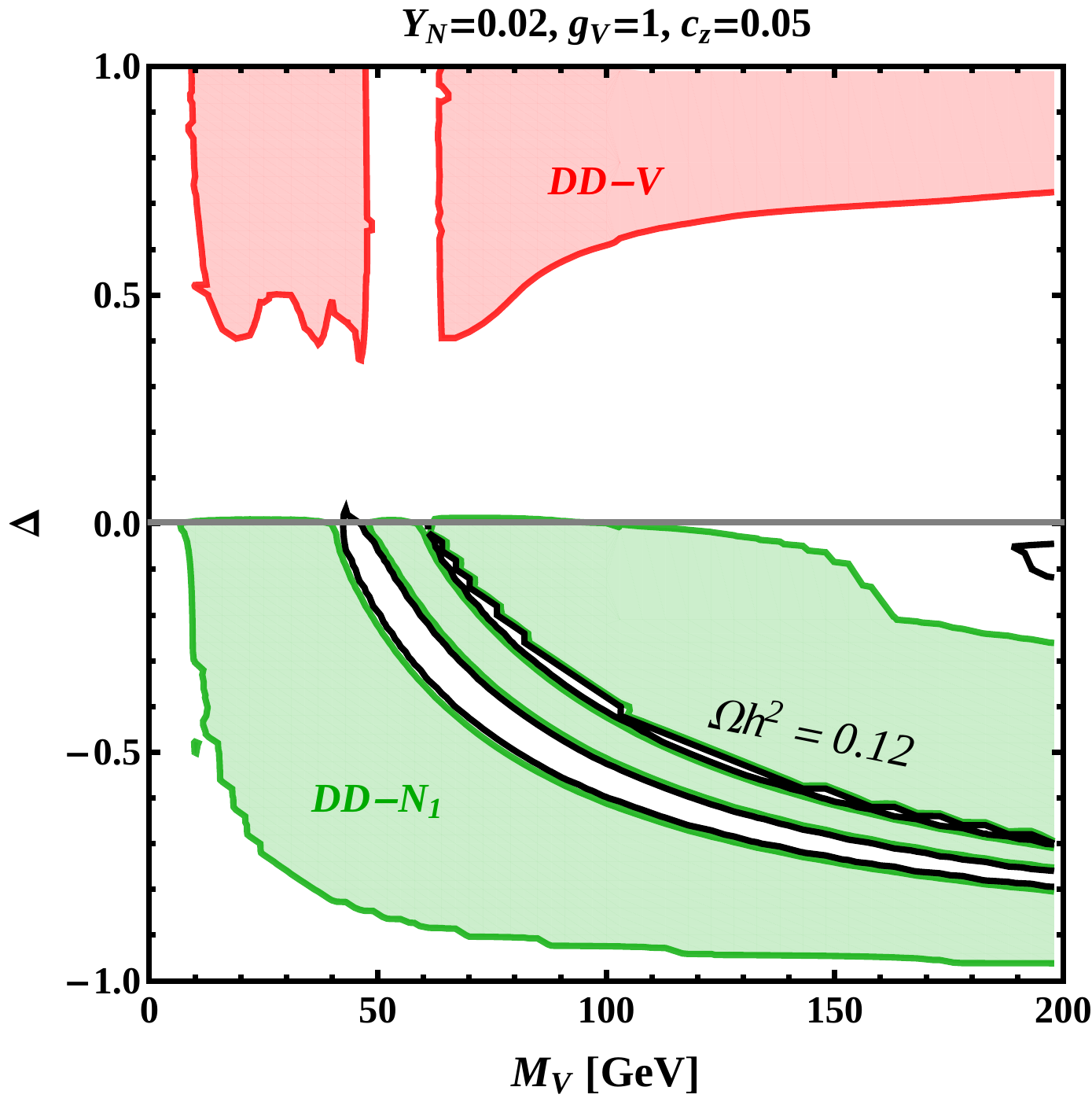}
\end{center}
\caption{Constraints from relic abundance, Invisible Higgs limits, and direct detection. Along the solid black curves, we have a relic abundance in agreement with the observed cosmological dark matter density. The blue shaded region indicates the limits from the invisible Higgs searches. The red and green shaded regions are excluded by the direct detection of the vector and fermion, respectively.}
\label{fig:successlim}
\end{figure}

There are two ways to reduce the fermion relic density without increasing the direct detection cross-section. The first is to increase $g_V$, the $U(1)^\prime$ gauge coupling. Semi-annihilation rates will increase, and will be most effective for $\Delta$ near zero. For positive $\Delta$, this also increases the rate that fermions convert into the vector candidate. However, if the fermion is too heavy to be a dynamical participant for the freeze-out of the vector DM, it suppresses the vector annihilation as it runs in the $h$-$V$-$V$ loop. Therefore, the large gauge coupling is still necessary to compensate this suppression and allow for efficient vector annihilation. 

The left plot of Fig.~\ref{fig:successlim} shows that this scenario can satisfy direct detection constraints and form a thermal relic. For these parameters, this is successful for $M_V \gtrsim 110$ GeV and $0.05 \lesssim \Delta \lesssim 0.15$, where $N_1$ makes up at most 1\% of the total DM abundance. Resonant vector annihilation through the Higgs, \eg $M_V \sim M_h/2$, with $\Delta\gtrsim 0.1$ can also avoid constraints. However, note that for larger values of $\Delta$, both $g_V$ and $Y_N$ may be safely increased and can allow for a sufficiently small relic abundance for the vector candidate.

The second promising avenue is to consider new annihilation channels induced by couplings to the $Z$. Since the $Z$ only couples directly to the fermion, this effect will be most relevant when the fermion is comparable in mass or lighter than the vector. Similar to the previous case, increasing $c_z$ increases fermion--fermion annihilation as well as semi-annihilation processes such as $NV\rightarrow NZ$ and $N\bar{N}\rightarrow VZ$. A small $Y_N$ will also be necessary to avoid direct detection constraints on the fermion. Further, if $Y_N$ is small enough, the invisible Higgs constraints can be evaded which allows for fermions light enough to resonantly annihilate through the $Z$.

The right plot in Fig.~\ref{fig:successlim} shows the available parameter space. In this case, the yukawa is small enough such that Higgs invisible constraints are not relevant in this plane. The small yukawa also severely reduces direct detection constraints. This set of parameters is viable for negative $\Delta$, with $M_{N_1}\sim M_Z/2$ or $M_h/2$, where the fermion resonantly annihilates through an s-channel $Z$ or Higgs, respectively. In this region, $N_1$ can make up all of the DM relic abundance, while $V$ can at most make up 10\% near $\Delta=0$.

This is also viable for $M_V\gtrsim 190$ GeV with $-0.15 \lesssim \Delta \lesssim -0.05$, where specifically $ZZ$ and $VZ$ final states allow for efficient annihilation. Here, $V$ makes up roughly 10\% of the relic, with $N_1$ making up 50--90\%. This latter window is not accessible to $Z$ decays and therefore may be further opened by increasing $c_z$.

\section{Summary and Conclusions}
\label{sec:summary}
In this work, we further investigate the radiative Higgs portal that was introduced in \cite{DiFranzo:2015nli}. We consider a simplified model where a vector and fermion arise from a dark sector $U(1)^{\prime}$, which are both stabilized by an imposed dark charge conjugation symmetry. The vector can annihilate through a Higgs portal at the radiative level, with the fermion running in the loop and the fermion can annihilate at tree level through the Higgs. This model further generates semi-annihilation channels for the DM candidates. We investigate the phenomenology of this model considering relic abundance, direct detection, and Invisible Higgs constraints.

This model is highly constrained by direct detection of the fermionic candidate, requiring that it compose a small fraction of the total relic abundance or having a small coupling to the Higgs. Constraints may be avoided by decreasing the yukawa coupling while increasing the $U(1)^\prime$ gauge coupling. The increased gauge coupling helps to enhance semi-annihilation processes, conversion of $N_1$ into $V$, and $V$ annihilation; without increasing the $N_1$--nucleon scattering cross-section.

The fermion coupling to the $Z$ may also be used to enhance new semi-annihilation channels as well as annihilation through the $Z$ resonance and $ZZ$ final state processes. Since the $Z$ only couples axially to $N_1$, contributions to direct detection are spin-dependent or velocity suppressed, therefore increasing this coupling may be done with little recourse from direct detection. Invisible $Z$ width constraints are also easily evaded.

\bigskip
\bigskip

\textbf{Acknowledgments.} We would like to thank Paddy Fox, Tim Tait, KC Kong, and David Shih for valuable discussions and direction. AD is supported in part by the Universities Research Association Visiting Scholars Award Program at Fermilab and by the NSF Grant No.~PHY-1316792. During the beginning of this work GM was supported by the Fermilab Graduate Student Research Program in Theoretical Physics and is now supported in part by the National Research Foundation of South Africa, Grant No. 88614 and by the University of Kansas, Physics and Astronomy dissertation fellowship. Fermilab is operated by Fermi Research Alliance, LLC, under Contract No. DE-AC02-07CH11359 with the US Department of Energy.

\appendix

\section{Loop functions for Relic density and Invisible Higgs width}
\label{app:appA}
\begin{align}
A_{inv}(M_h, M_{N_1}, M_V) = \frac{M_{N_1}}{2 \sqrt{2}\pi^2} \bigg(4 C_{12} - C_0 \bigg)
\end{align}
\begin{align}
B_{inv}(M_h, M_{N_1}, M_V) = \frac{M_{N_1} }{2 \sqrt{2}\pi^2} \Bigg[\frac{1}{2} + M_{N_1}^2~C_0 - M_V^2 \bigg(C_{11} + C_{22} \bigg) + \bigg(2 M_V^2 - M_h^2\bigg)~C_{12} \Bigg]
\end{align}
\noindent Here $C_0$ and $C_{ij}$ are the Passarino--Veltman coefficients as defined in FormCalc and LoopTools \cite{Hahn:1998yk}. All $C_{X}$'s here are evaluated as: $C_X[M_V^2, M_h^2, M_V^2, M_{N_1}^2, M_{N_1}^2, M_{N_1}^2]$.

\section{Loop functions for Direct Detection}
\label{app:appB}

The loop functions for the direct detection rates are calculated as follows: 
\begin{align}
A_{DD}(t, M_{N_1}, M_V) = \frac{M_{N_1}}{2 \sqrt{2}\pi^2} \bigg(4 C_{12} - C_0 \bigg)
\end{align}
and 
\begin{align}
B_{DD}(t, M_{N_1}, M_V) = \frac{M_{N_1}}{2 \sqrt{2} \pi^2} \Bigg[B_0[t, M_{N_1}^2, M_{N_1}^2] - 4 ~C_{00} + 4 \bigg(\frac{t}{2} - M_V^2 \bigg) C_{12} \Bigg].
\end{align}
\noindent Here $B_0$, $C_0$, and $C_{ij}$ are the Passarino--Veltman coefficients as defined in FormCalc and LoopTools \cite{Hahn:1998yk}. All $C_{X}$'s here are evaluated as: $C_X[M_V^2, t, M_V^2, M_{N_1}^2, M_{N_1}^2, M_{N_1}^2]$.

We define $F_0$ which is a function of the vector and the fermion mass and uses the zero momentum transfer approximation for dark matter scattering through the Higgs. 
\begin{equation}
F_0(M_{N_1}, M_V) = \lim_{t\rightarrow 0^-} \big[B_{DD}(t, M_{N_1}, M_V) - M_V^2 \, A_{DD}(t, M_{N_1}, M_V)\big],
\end{equation}

\bibliography{HP}
\bibliographystyle{JHEP}

\end{document}